\documentclass[journal]{IEEEtran}\newcommand{\figsize}{3.4}
\usepackage{graphicx,psfrag}
\usepackage[usenames]{color}
\usepackage[cmex10]{amsmath}
\usepackage{amsfonts,amssymb,latexsym,cite,color}

 \newcommand{\putFrag}[4]{\begin{figure}[ht]
                            \centering
                            #4
			    \includegraphics[width=#3in]{figures/#1.eps}
            		    \caption{#2}
           		    \label{fig:#1}
                          \end{figure} }

 \newcommand{\defn}{\triangleq}

 \newcommand{\tvec}[1]{\ensuremath{\Tilde{\boldsymbol{#1}}}}
 
 \newcommand{\hvec}[1]{\ensuremath{\Hat{\boldsymbol{#1}}}}
    
 \renewcommand{\vec}[1]{\ensuremath{\boldsymbol{#1}}}

 \newcommand{\mc}[1]{\ensuremath{\mathcal{#1}}}
 
 \newcommand{\barst}[1]{\ensuremath{\text{\raisebox{-0.5mm}{$\bigl|_{#1}$}}}}

 \newcommand{\Complex}{{\mathbb{C}}}
 \newcommand{\Int}{{\mathbb{Z}}}

 \newcommand{\tran}{^{\textsf{T}}}
 \newcommand{\herm}{^{\textsf{H}}}
 \newcommand{\of}[1]{^{(#1)}}

 \newcommand{\ofH}[1]{^{{(#1)}\textsf{H}}}

 \DeclareMathOperator{\E}{E}
 
 \DeclareMathOperator{\cov}{Cov}

 \DeclareMathOperator{\tr}{tr}
 \DeclareMathOperator{\diag}{diag}

 \DeclareMathOperator{\vect}{vec}
 
 \DeclareMathOperator{\Diag}{Diag}

 \DeclareMathOperator{\Perp}{\perp\!\!\!\perp}
 \DeclareMathOperator{\elsum}{sum}

 \renewcommand{\eqref}[1]{(\ref{eq:#1})}
 
 \newcommand{\Figref}[1]{Figure~\ref{fig:#1}}
 \newcommand{\figref}[1]{Fig.~\ref{fig:#1}}
 
 \newcommand{\secref}[1]{Section~\ref{sec:#1}}
 
 \newcommand{\appref}[1]{Appendix~\ref{app:#1}}

 \newcounter{comment}[section]
 
 \newcounter{texthead}[section]

 \newcommand{\IEEElabel}[1]{\begingroup\addtocounter{equation}{-1}
 	\refstepcounter{equation}\label{#1}\endgroup}

\newcommand{\giv}{\,|\,}
\newcommand{\biggiv}{\,\big|\,}
\newcommand{\Biggiv}{\,\Big|\,}

\newcommand{\s}{_\textsf{s}}
\newcommand{\R}{_\textsf{r}}
\newcommand{\D}{_\textsf{d}}
\newcommand{\sR}{_\textsf{sr}}
\newcommand{\sD}{_\textsf{sd}}
\newcommand{\RR}{_\textsf{rr}}
\newcommand{\RD}{_\textsf{rd}}

\newcommand{\sRt}{_{\textsf{sr},\tau}}
\newcommand{\RDt}{_{\textsf{rd},\tau}}

\newcommand{\Ns}{N\s} 
\newcommand{\NR}{N\R} 
\newcommand{\MR}{M\R}
\newcommand{\MD}{M\D}

\newcommand{\fd}{_{\textsf{FD}}}
\newcommand{\hd}{_{\textsf{HD}}}
\newcommand{\train}{_\textsf{train}}
\newcommand{\data}{_\textsf{data}}
\newcommand{\dB}{\text{dB}}
\newcommand{\QQ}{\vec{\mc{Q}}}
\newcommand{\QQoptt}{\vec{\mc{Q}}_{*,\tau}} 

\newcommand{\Qcont}{\mathbb{Q}_{\tau}} 
\newcommand{\Qopt}{\mathbb{Q}_*}
\newcommand{\Qoptt}{\mathbb{Q}_{*,\tau}} 

\newcommand{\Qeqt}{\mathbb{Q}_{=,\tau}} 
\newcommand{\crit}{_\textsf{crit}}
\newcommand{\zeq}{\zeta_=} 
\newcommand{\Iopt}{\underline{I}_*}
\newcommand{\Ioptt}{\underline{I}_{*,\tau}} 

\begin{document}
\setlength{\arraycolsep}{0.8mm}

 \title{Full-Duplex MIMO Relaying: Achievable Rates under Limited Dynamic Range}
 	\author{Brian P. Day,\IEEEauthorrefmark{1}
		Adam R. Margetts,\IEEEauthorrefmark{2} 
 		Daniel W. Bliss,\IEEEauthorrefmark{2} 
		and Philip Schniter\IEEEauthorrefmark{1}\IEEEauthorrefmark{3}
	\thanks{\IEEEauthorrefmark{1}Brian Day and Philip Schniter are with the Department of Electrical and Computer Engineering, The Ohio State University, Columbus, OH.}%
	\thanks{\IEEEauthorrefmark{2}Daniel Bliss and Adam Margetts are with the Advanced Sensor Techniques Group, MIT Lincoln Laboratory, Lexington, MA.}%
	\thanks{\IEEEauthorrefmark{3}Please direct all correspondence to Prof. Philip Schniter, Dept. ECE, The Ohio State University, 2015 Neil Ave., Columbus OH 43210, e-mail: schniter@ece.osu.edu, phone 614.247.6488, fax 614.292.7596.}
	\thanks{Manuscript received August 25, 2011; revised May 14, 2012.}%
	\thanks{This work was sponsored by the Defense Advanced Research Projects Agency under Air Force contract FA8721-05-C-0002.  Opinions, interpretations, conclusions, and recommendations are those of the authors and are not necessarily endorsed by the United States Government.}%
	}
 \date{\today}
 \maketitle

\begin{abstract}
In this paper we consider the problem of full-duplex multiple-input multiple-output 
(MIMO) relaying between multi-antenna source and destination nodes.
The principal difficulty in implementing such a system is that, due to the limited 
attenuation between the relay's transmit and receive antenna arrays, the relay's 
outgoing signal may overwhelm its limited-dynamic-range input circuitry, 
making it difficult---if not impossible---to recover the desired incoming signal.
While explicitly modeling transmitter/receiver dynamic-range limitations 
and channel estimation error,
we derive tight upper and lower bounds on the end-to-end achievable rate 
of decode-and-forward-based full-duplex MIMO relay systems, and 
propose a transmission scheme based on maximization of the lower bound.
The maximization requires us to (numerically) solve a nonconvex optimization 
problem, for which we detail a novel approach based on bisection search and 
gradient projection.
To gain insights into system design tradeoffs, we also derive an analytic 
approximation to the achievable rate and numerically demonstrate its accuracy.
We then study the behavior of the achievable rate as a function
of signal-to-noise ratio, interference-to-noise ratio, transmitter/receiver 
dynamic range, number of antennas, and training length, 
using optimized half-duplex signaling as a baseline.


\end{abstract}

\section{Introduction}			\label{sec:intro}

 We consider the problem of communicating from a source node to a destination
 node through a relay node.
 Traditional relay systems operate in a half-duplex mode, whereby the time-frequency
 signal-space used for the source-to-relay link is kept orthogonal to that 
 used for the relay-to-destination link, such as with non-overlapping time periods
 or frequency bands.
 Half-duplex operation is used to avoid the high levels of relay self-interference 
 that are faced with full-duplex\footnote{Successful full-duplex communication has 
 been recently demonstrated in the non-relay setting \cite{Jain:MOBICOM:11} and in 
 the non-MIMO relay setting \cite{Everett:ASIL:11}. }
 operation (see \figref{relay_phy}), where
 the source and relay share a common time-frequency signal-space.
 For example, it is not unusual for the ratio between the
 relay's self-interference power and desired incoming signal power to exceed 100~dB 
 \cite{Hua:MIL:10}, or---in general---some value larger than the dynamic range of the relay's 
 front-end hardware, making it impossible to recover the desired signal. 
 The importance of \emph{limited dynamic-range} (DR) cannot be overstressed; 
 notice that, even if the self-interference signal was perfectly known, limited-DR renders 
 perfect cancellation impossible.
\putFrag{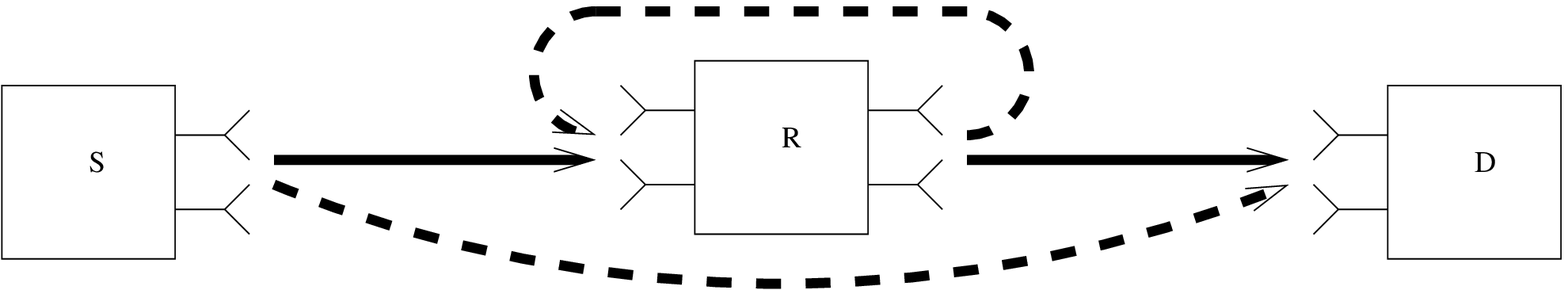}
	{Full-duplex MIMO relaying from source to destination.
	 Solid lines denote desired propagation and dashed
	 lines denote interference.}
	{2.25}
        {\psfrag{S}[][Bl][1]{\sf S}
         \psfrag{R}[][Bl][1]{\sf R}
         \psfrag{D}[][Bl][1]{\sf D}}

 Recently, multiple-input multiple-output (MIMO) relaying has been proposed 
 as a means of increasing spectral efficiency (e.g., 
 \cite{
 Wang:TIT:05,
 Simoens:TSP:09
 }).
 By MIMO relaying, we mean that the source, relay, and destination each use
 multiple antennas for both reception and transmission.
 MIMO relaying brings the possibility of full-duplex operation
 through \emph{spatial} self-interference suppression (e.g., 
 \cite{
 Bliss:SSP:07,
 PLarsson:VTC:09,
 Riihonen:ASIL:09,
 Hua:MIL:10,
 Riihonen:ASIL:10,
 Riihonen:CISS:11,
 Sohaib:VTC:09,
 Sangiamwong:VTC:09,
 BChun:APSIPA:09, 
 BChun:WCNC:10,
 Lioliou:ASIL:10 
 }).
 As a simple example, one can imagine using the relay's 
 transmit array to form spatial nulls at a subset of the relay's receive antennas, which 
 are then free of self-interference and able to recover the desired signal.
 In forming these nulls, however, it can be seen that the relay consumes spatial 
 degrees-of-freedom that could have been used in communicating data to the destination.
 Thus, maximizing the end-to-end throughput involves navigating a tradeoff between the 
 source-to-relay link and relay-to-destination link.
 Of course, maximizing end-to-end throughput is more involved than simply protecting an
 arbitrary subset of the relay's receive antennas; 
 one also needs to consider which subset to protect, and the degree to which each of
 those antennas are protected, given the source-to-relay and relay-to-destination MIMO channel
 coefficients, the estimation errors on those coefficients, and the DR limitations of 
 the various nodes.
 These considerations motivate the following fundamental questions about full-duplex 
 MIMO relaying in the presence of self-interference: 
 \emph{1) What is the maximum achievable end-to-end throughput under a transmit power constraint? 
 2) How can the system be designed to achieve this throughput?}
 
 In this paper, we aim to answer these two fundamental questions while paying special
 attention to the effects of both limited-DR and channel estimation error. 
 \begin{enumerate}
 \item
 Limited-DR is a natural consequence of non-ideal amplifiers, oscillators, 
 analog-to-digital converters (ADCs),
 and digital-to-analog converters (DACs). 
 To model the effects of limited receiver-DR, we inject, at each receive antenna, 
 an additive white Gaussian ``receiver distortion'' with variance $\beta$ 
 times the energy impinging on that receive antenna (where $\beta\ll 1$).
 Similarly, 
 to model the effects of limited transmitter-DR, we inject, at each transmit 
 antenna, an additive white Gaussian ``transmitter noise'' with variance 
 $\kappa$ times the energy of the intended transmit signal (where $\kappa\ll 1$).
 Thus, $\kappa^{-1}$ and $\beta^{-1}$ characterize the transmitter 
 and receiver dynamic ranges, respectively.
 \item
 Imperfect CSI can result for several reasons, including 
 channel time-variation, additive noise, and DR limitations.
 We focus on CSI imperfections that result from the use of pilot-aided 
 least-squares (LS) channel estimation performed in the presence of limited-DR.
 \end{enumerate}
 Moreover, we consider regenerative relays that decode-and-forward 
 (as in \cite{
 Bliss:SSP:07,
 PLarsson:VTC:09,
 Riihonen:ASIL:09,
 Riihonen:ASIL:10,
 Hua:MIL:10,
 Riihonen:CISS:11
 }),
 as opposed to simpler non-regenerative relays that only amplify-and-forward
 (as in \cite{
 Sohaib:VTC:09,
 Sangiamwong:VTC:09,
 BChun:APSIPA:09, 
 BChun:WCNC:10,
 Lioliou:ASIL:10 
 }).

The contributions of this paper are as follows.
For the full-duplex MIMO relaying problem, 
an explicit model for transmitter/receiver-DR limitations is proposed;
pilot-aided least-squares MIMO-channel estimation, under DR limitations, is analyzed;
the residual self-interference, from DR limitations and channel-estimation error, is analyzed; 
lower and upper bounds on the achievable rate are derived; 
a transmission scheme is proposed based on maximizing the achievable-rate lower bound subject to a power constraint, requiring the solution of a nonconvex optimization problem, to which we apply bisection search and Gradient Projection;
an analytic approximation of the maximum achievable rate is proposed; and,
the achievable rate is numerically investigated as a function of signal-to-noise ratio, interference-to-noise ratio, transmitter/receiver dynamic range, number of antennas, and number of pilots. 

The paper is structured as follows.
In \secref{model}, we state our channel model, limited-DR model, and assumptions on the transmission protocol.
Then, in \secref{analysis}, we derive upper and lower bounds on the achievable rate under pilot-aided channel estimation and partial self-interference cancellation at the relay.
In \secref{approach}, we propose a novel transmission scheme that is based on 
maximizing the achievable-rate lower-bound subject to a power constraint and,
in \secref{approx}, we derive a closed-form approximation of the optimized achievable rate whose accuracy is numerically verified.
Then, in \secref{sims}, we numerically investigate achievable rate as a function of the SNRs $(\rho\R,\rho\D)$, the INRs $(\eta\R,\eta\D)$, the dynamic range parameters $(\kappa,\beta)$, the number of antennas $(N\R,N\D)$, and the training length $T$, and we also investigate the gain of full-duplex signaling (over half-duplex) and partial self-interference cancellation.
Finally, in \secref{conclusion}, we conclude.


\emph{Notation}:
  We use 
  $(\cdot)\tran$ to denote transpose, 
  $(\cdot)^*$ conjugate, and 
  $(\cdot)\herm$ conjugate transpose.
  For matrices $\vec{A},\vec{B}\in\Complex^{M\times N}$, we use
  $\tr(\vec{A})$ to denote trace, 
  $\det(\vec{A})$ to denote determinant, 
  $\vec{A}\odot\vec{B}$ to denote elementwise (i.e., Hadamard) product, 
  $\elsum(\vec{A})\in\Complex$ to denote the sum over all elements,
  $\vect(\vec{A})\in\Complex^{MN}$ to denote vectorization,
  $\diag(\vec{A})$ to denote the diagonal matrix with the same diagonal 
  	elements as $\vec{A}$, 
  $\Diag(\vec{a})$ to denote the diagonal matrix whose diagonal is constructed 
  	from the vector $\vec{a}$,
  and
  $[\vec{A}]_{m,n}$ to denote the element in the $m^{th}$ row 
  	and $n^{th}$ column of $\vec{A}$. 
  We denote expectation by $\E\{\cdot\}$,
  covariance by $\cov\{\cdot\}$,
  statistical independence by $\Perp$,
  the circular complex Gaussian pdf with mean vector $\vec{m}$ and 
  covariance matrix $\vec{Q}$ by $\mc{CN}(\vec{m},\vec{Q})$,
  and
  the Kronecker delta sequence by $\delta_k$.
  Finally,
  $\vec{I}$ denotes the identity matrix,
  $\Complex$ the complex field, and
  $\Int^+$ the positive integers. 

\section{ System Model }				\label{sec:model}
We will use $\Ns$ and $\NR$ to denote the number of transmit antennas at the
source and relay, respectively, and $\MR$ and $\MD$ to denote the 
number of receive antennas at the relay and destination, respectively.
Here and in the sequel, we use 
subscript-\textsf{s} for source,
subscript-\textsf{r} for relay, and
subscript-\textsf{d} for destination. 
Similarly, we will use
subscript-\textsf{sr} for source-to-relay,
subscript-\textsf{rd} for relay-to-destination, 
subscript-\textsf{rr} for relay-to-relay, and
subscript-\textsf{sd} for source-to-destination.
At times, we will omit the subscripts when referring to common quantities.
For example, we will use 
$\vec{s}(t)\in\Complex^{N}$ to denote the time $t\!\in\!\Int^+$ noisy signals 
radiated by the transmit antenna arrays, and 
$\vec{u}(t)\in\Complex^{M}$ to denote the time-$t$ undistorted signals 
collected by the receive antenna arrays. 
More specifically, the source's and relay's radiated signals are 
$\vec{s}\s(t)\in\Complex^{\Ns}$ and $\vec{s}\R(t)\in\Complex^{\NR}$, respectively, 
while the relay's and destination's collected signals are 
$\vec{u}\R(t)\in\Complex^{\MR}$ and $\vec{u}\D(t)\in\Complex^{\MD}$, respectively. 

\subsection{ Propagation Channels }			\label{sec:chan}

We assume that propagation between each transmitter-receiver pair 
can be characterized by 
a Raleigh-fading MIMO channel $\vec{H}\in\Complex^{M\times N}$ 
corrupted by additive white Gaussian noise (AWGN) $\vec{n}(t)$.
By ``Rayleigh fading,'' we mean that 
$\vect(\vec{H})\sim\mc{CN}(\vec{0},\vec{I}_{MN})$,
and by ``AWGN,'' we mean that 
$\vec{n}(t) \sim \mc{CN}(\vec{0},\vec{I}_{M})$.
The time-$t$ radiated signals $\vec{s}(t)$ are then related 
to the received signals $\vec{u}(t)$ via
\begin{align}
  \vec{u}\R(t)
  &= \sqrt{\rho\R} \vec{H}\sR \vec{s}\s(t) 
	+ \sqrt{\eta\R} \vec{H}\RR\vec{s}\R(t) + \vec{n}\R(t) 
							\label{eq:uR}\\
 \vec{u}\D(t)
 &= \sqrt{\rho\D} \vec{H}\RD \vec{s}\R(t) 
	+ \sqrt{\eta\D} \vec{H}\sD\vec{s}\s(t) + \vec{n}\D(t) .
							\label{eq:uD}
\end{align}
In \eqref{uR}-\eqref{uD}, 
$\rho\R>0$ and $\rho\D>0$ denote the signal-to-noise ratio (SNR) at the
relay and destination, while
$\eta\R>0$ and $\eta\D>0$ denote the interference-to-noise ratio (INR) 
at the relay and destination. 
(As described in the sequel, the destination treats the source-to-destination 
link as interference).
The INR $\eta\R$ will depend on the separation between, and 
orientation of, the relay's transmit and receive antenna arrays
\cite{Riihonen:CISS:11}, whereas
the INR $\eta\D$ will depend on the separation between source and
destination modems, so that typically $\eta\D \ll \eta\R$.
We emphasize that \eqref{uR}-\eqref{uD} models the channels $\vec{H}\sR$,
$\vec{H}\RR$, $\vec{H}\RD$, and $\vec{H}\sD$, as time-invariant quantities.

\subsection{ Transmission Protocol }		\label{sec:protocol}

For full-duplex decode-and-forward relaying, we partition the time indices
$t=0,1,2,\dots$ into a sequence of communication epochs 
$\{\mc{T}_i\}_{i=0}^\infty$ where, during epoch $\mc{T}_i\subset\Int^+$,
the source communicates the $i^{th}$ information packet to the relay, while 
simultaneously the relay communicates the $(i\!-\!1)^{th}$ information packet 
to the destination.  
Before the first data communication epoch, we assume the existence of a 
training epoch $\mc{T}\train$ during which the modems estimate the 
channel state. 
From the estimated channel state, the data communication design parameters are 
optimized and the resulting parameters are used for every data communication epoch.
Since the design and analysis will be identical for every data-communication 
epoch (as a consequence of channel time-invariance), 
we suppress the index $i$ in the sequel and refer to an arbitrary 
data communication epoch as $\mc{T}\data$.

The training epoch is partitioned into two equal-length periods 
(i.e., $\mc{T}\train[1]$ and $\mc{T}\train[2]$) to avoid 
self-interference when estimating the channel matrices. 
Each data epoch is also partitioned into two periods 
(i.e., $\mc{T}\data[1]$ and $\mc{T}\data[2]$) of normalized duration $\tau\in[0,1]$ and $1-\tau$,
respectively, over which the transmission parameters can be independently optimized.
As we shall see in the sequel, such flexibility is critical when the INR $\eta\R$ is large relative to the SNR $\rho\R$.
Moreover, this latter partitioning allows us to formulate both half- and full-duplex 
schemes as special cases of a more general transmission protocol.
For use in the sequel, we find it convenient to define $\tau[1] \defn \tau$ and $\tau[2]\defn1-\tau$.
Within each of these periods, we assume that the 
transmitted signals are zero-mean and wide-sense stationary.

\subsection{ Limited Transmitter Dynamic Range }	\label{sec:lim_tdr}

We model the effect of limited transmitter dynamic range (DR) by injecting, 
per transmit antenna, an independent zero-mean Gaussian ``transmitter noise'' 
whose variance is $\kappa$ times the energy of the \emph{intended} transmit 
signal at that antenna. 
In particular, say that $\vec{x}(t)\in\Complex^{N}$ denotes 
the transmitter's intended time-$t$ transmit signal, and say 
$\vec{Q}\defn\cov\{\vec{x}(t)\}$ over the relevant time period 
(e.g., $t\in\mc{T}\data[1]$). 
We then write the time-$t$ noisy radiated signal as
\begin{equation}
 \vec{s}(t) 
 = \vec{x}(t) + \vec{c}(t)
 ~ \text{s.t.} 
 \left\{
 \begin{array}{l}
   \vec{c}(t) \sim \mc{C}\mc{N}(\vec{0},\kappa \diag(\vec{Q})) \\
   \vec{c}(t) \Perp \vec{x}(t)\\
   \vec{c}(t) \Perp \vec{c}(t')\barst{t'\neq t}    \quad ,
 \end{array}
 \right.
 						\label{eq:tx}
\end{equation}
where $\vec{c}(t)\in\Complex^{N}$ denotes transmitter noise
and $\Perp$ statistical independence. 
Typically, $\kappa\ll 1$.
As shown by measurements of various hardware setups (e.g., \cite{Santella:TVT:98,Suzuki:JSAC:08}), the independent Gaussian noise model in
\eqref{tx} closely approximates the combined effects of
additive power-amp noise, non-linearities in the DAC and power-amp,
and oscillator phase noise.
Moreover, the dependence of the transmitter-noise variance on intended 
signal power in \eqref{tx} follows directly from the definition of 
limited dynamic range.

\subsection{ Limited Receiver Dynamic Range }		\label{sec:lim_rdr}

We model the effect of limited receiver-DR by injecting,
per receive antenna, an independent zero-mean Gaussian ``receiver 
distortion'' whose variance is $\beta$ times the energy collected 
by that antenna.
In particular, say that $\vec{u}(t)\in\Complex^{M}$ denotes the 
receiver's undistorted time-$t$ received vector, and say
$\vec{\Phi}\defn\cov\{\vec{u}(t)\}$
over the relevant time period (e.g., $t\in\mc{T}\data[1]$). 
We then write the distorted post-ADC received signal as
\begin{equation}
 \vec{y}(t) 
 = \vec{u}(t) + \vec{e}(t)
 ~ \text{s.t.} ~
 \left\{
 \begin{array}{l}
   \vec{e}(t) \sim \mc{C}\mc{N}(\vec{0},\beta \diag(\vec{\Phi})) \\
   \vec{e}(t) \Perp \vec{u}(t)\\
   \vec{e}(t) \Perp \vec{e}(t')\barst{t'\neq t}   \quad ,
 \end{array}
 \right.
 							\label{eq:rx}
\end{equation}
where $\vec{e}(t)\in\Complex^{M}$ is additive distortion. 
Typically, $\beta\ll 1$.
From a theoretical perspective, automatic gain control (AGC) followed by dithered uniform quantization \cite{Gray:TIT:93} yields quantization errors whose statistics closely match the model \eqref{rx}.
More importantly, studies (e.g., \cite{Namgoong:TWC:05}) have shown that the independent Gaussian distortion model \eqref{rx} accurately captures the combined effects of additive AGC noise, non-linearities in the ADC and gain-control, and oscillator phase noise in practical hardware.

\Figref{relay_direct} summarizes our model.  
The dashed lines indicate that the distortion levels are proportional to mean energy levels and not to the instantaneous value.

\putFrag{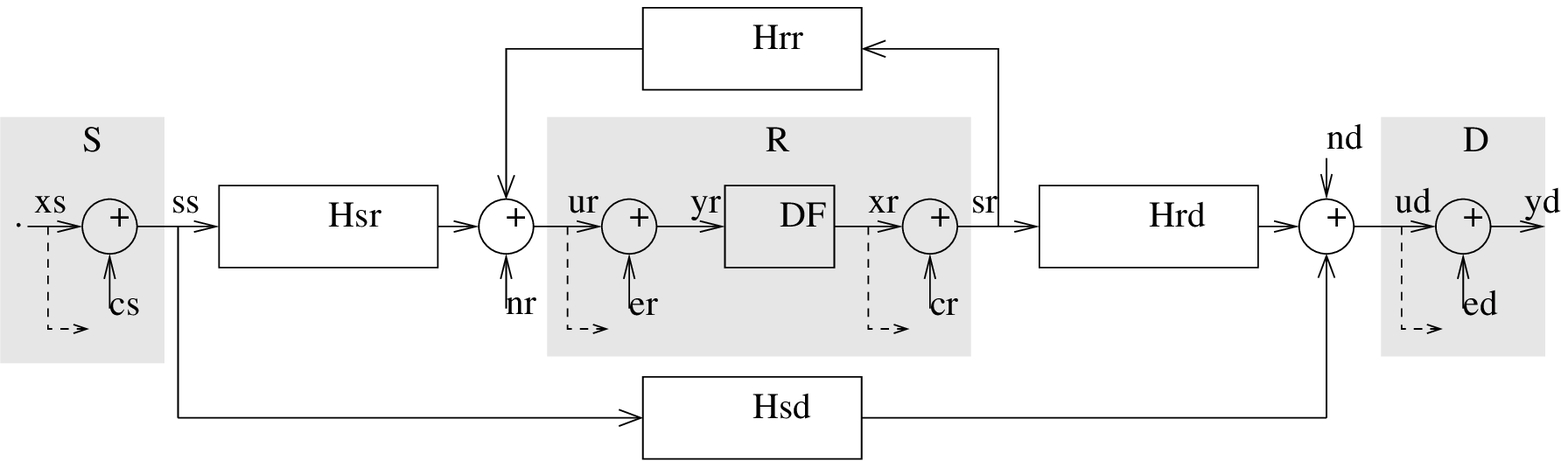}
	{Our model of full-duplex MIMO relaying 
 	 under limited transmitter/receiver-DR. 
	 The dashed lines denote statistical dependence.}
	{\figsize}
        {\newcommand{\sz}{0.9}
	 \psfrag{.}{}
	 \psfrag{+}[][Bl][\sz]{$+$}
         \psfrag{Hsr}[][Bl][\sz]{$\sqrt{\rho\R}\vec{H}\sR$}
	 \psfrag{Hrr}[][Bl][\sz]{$\sqrt{\eta\R}\vec{H}\RR$}
	 \psfrag{Hrd}[][Bl][\sz]{$\sqrt{\rho\D}\vec{H}\RD$}
	 \psfrag{Hsd}[][Bl][\sz]{$\sqrt{\eta\D}\vec{H}\sD$}
	 \psfrag{xs}[b][Bl][\sz]{$\vec{x}\s$}
	 \psfrag{cs}[t][Bl][\sz]{$\vec{c}\s$}
	 \psfrag{ss}[b][Bl][\sz]{$\vec{s}\s$}
	 \psfrag{nr}[t][Bl][\sz]{$\vec{n}\R$}
	 \psfrag{ur}[b][Bl][\sz]{$\vec{u}\R$}
	 \psfrag{er}[t][Bl][\sz]{$\vec{e}\R$}
	 \psfrag{yr}[b][Bl][\sz]{$\vec{y}\R$}
	 \psfrag{DF}[][Bl][0.37]{\begin{tabular}{c}\textsf{decode \&}\\\textsf{forward}\end{tabular}}
	 \psfrag{S}[][Bl][0.65]{\textsf{Source}}
	 \psfrag{R}[][Bl][0.7]{\textsf{Relay}}
	 \psfrag{D}[][Bl][0.7]{\textsf{Dest}}
	 \psfrag{xr}[b][Bl][\sz]{$\vec{x}\R$}
	 \psfrag{cr}[t][Bl][\sz]{$\vec{c}\R$}
	 \psfrag{sr}[b][Bl][\sz]{$\vec{s}\R$}
	 \psfrag{nd}[b][Bl][\sz]{$\vec{n}\D$}
	 \psfrag{ud}[b][Bl][\sz]{$\vec{u}\D$}
	 \psfrag{ed}[t][Bl][\sz]{$\vec{e}\D$}
	 \psfrag{yd}[b][Bl][\sz]{$\vec{y}\D$}
	 }


\section{ Analysis of Achievable Rate }		\label{sec:analysis}

\subsection{ Pilot-Aided Channel Estimation }		\label{sec:chan_est}

In this section, we describe the pilot-aided channel estimation procedure 
that is used to learn the channel matrices $\vec{H}$.
In our protocol, the training epoch consists of two periods, 
$\mc{T}\train[1]$ and $\mc{T}\train[2]$, each spanning
$T N$ channel uses (for some $T\in\Int^+$). 
For all times $t\in\mc{T}\train[1]$, we assume that the source transmits 
a known pilot signal and the relay remains silent, while, for all 
$t\in\mc{T}\train[2]$, the relay transmits and the source remains silent.
Moreover, we construct the pilot sequence
$\vec{X} = [\vec{x}(1),\dots,\vec{x}(TN)]\in\Complex^{N\times TN}$ 
to satisfy $\frac{1}{2T}\vec{X} \vec{X}\herm = \vec{I}_{N}$, 
where the scaling has been chosen to satisfy a per-period power constraint 
of the form $\tr(\vec{Q}) = 2$, consistent with the data power constraints 
that will be described in the sequel.

Our limited transmitter/receiver-DR model implies that the (distorted) 
space-time pilot signal observed by a given receiver takes the form 
\begin{align}
  \vec{Y}
  &= \sqrt{\alpha} \vec{H}(\vec{X} + \vec{C})
	+ \vec{N} + \vec{E} ,   			\label{eq:Y}
\end{align}
where $\alpha\in\{\rho\R,\eta\R,\rho\D, \eta\D\}$ for 
$\vec{H}\in\{\vec{H}\sR,\vec{H}\RR,\vec{H}\RD, \vec{H}\sD\}$, respectively.
In \eqref{Y}, $\vec{C}, \vec{E}$ and $\vec{N}$ are $N\times TN$ matrices 
of transmitter noise, receiver distortion, and AWGN, respectively.
At the conclusion of training, we assume that each receiver 
uses least-squares (LS) to estimate the corresponding channel $\vec{H}$ as
\begin{align}
  \sqrt{\alpha}\hvec{H}
  &\triangleq \frac{1}{2T} \vec{Y} \vec{X}\herm ,
							\label{eq:Hhat}
\end{align}
and communicates this estimate to the transmitter.\footnote{
  In our transmission protocol, a single training epoch is followed by
  a large number of data epochs, and so the relative training overhead becomes 
  negligible as the number of data epochs grows large.}
In the sequel, it will be useful to decompose the channel estimate into the 
true channel plus an estimation error.
In \appref{chan_est}, it is shown that such a decomposition takes the form 
\begin{align}
  \sqrt{\alpha}\hvec{H}
  &= \sqrt{\alpha} \vec{H}  +  \vec{D}^{\frac{1}{2}} \tvec{H},	
  							\label{eq:Htilde}
\end{align}
where the entries of $\tvec{H}$ are i.i.d $\mc{CN}(0,1)$, and where 
\begin{align}
  \vec{D}
  &=   \frac{1}{2T}
	\bigg(  (1 + \beta)\vec{I}  +  \alpha \frac{2\kappa}{N} \vec{H}\vec{H}\herm 
\nonumber\\&\quad
	 + \alpha \frac{2\beta}{N} (1 + \kappa) \diag\Big( \vec{H} \vec{H}\herm\Big) 
	\bigg)					\label{eq:chan_est_err}
\end{align}
characterizes the spatial covariance of the estimation error.  
Using $\beta \ll 1$ and $\kappa \ll 1$, this covariance reduces to
\begin{align}
 \vec{D}
 &\approx   \frac{1}{2T}
	\bigg(  \vec{I}  +  \alpha \frac{2\kappa}{N} \vec{H}\vec{H}\herm
	     + \alpha \frac{2\beta}{N} \diag\Big( \vec{H} \vec{H}\herm\Big) 
	\bigg) 	. 
\end{align}

\subsection{ Interference Cancellation and Equivalent Channel }   	\label{sec:cancellation}

We now describe how the relay partially cancels its self-interference,
and construct a simplified model for the result.

Recall that the data communication period is partitioned into two 
periods, $\mc{T}\data[1]$ and $\mc{T}\data[2]$,
and that---within each---the transmitted signals are wide-sense stationary.
Thus, at any time $t\in\mc{T}\data[l]$, the relay's (instantaneous, distorted) 
observed signal takes the form 
\begin{align}
   \vec{y}\R(t)
   &=  (\sqrt{\rho\R}\hvec{H}\sR - \vec{D}\sR^{\frac{1}{2}}\tvec{H}\sR)
		(\vec{x}\s(t) + \vec{c}\s(t))
	+ \vec{n}\R(t) + \vec{e}\R(t) 
   \nonumber\\&\quad
   	+ (\sqrt{\eta\R}\hvec{H}\RR - \vec{D}\RR^{\frac{1}{2}}\tvec{H}\RR)
		(\vec{x}\R(t) + \vec{c}\R(t)) ,
			\label{eq:y}
\end{align}
as implied by \figref{relay_direct} and \eqref{Htilde}.
Defining the aggregate noise term
\begin{align}
   \vec{v}\R(t)
	&\defn   \sqrt{\rho\R}\hvec{H}\sR\vec{c}\s(t)
	   -  \vec{D}^{\frac{1}{2}}\sR\tvec{H}\sR(\vec{x}\s(t) + \vec{c}\s(t))
	+  \vec{n}\R(t)  
\nonumber\\&\quad
	+   \vec{e}\R(t)  
	+  \sqrt{\eta\R}\hvec{H}\RR\vec{c}\R(t)
	-  \vec{D}\RR^{\frac{1}{2}}\tvec{H}\RR(\vec{x}\R(t) + \vec{c}\R(t)) ,
			\label{eq:v}
\end{align}
we can write the observed signal as
$\vec{y}\R(t) = \sqrt{\rho\R}\hvec{H}\sR\vec{x}\s(t)
	   + \sqrt{\eta\R}\hvec{H}\RR\vec{x}\R(t) +  \vec{v}\R(t)$,
where the self-interference term $\sqrt{\eta\R}\hvec{H}\RR\vec{x}\R(t)$ is
known and thus can be canceled. 
The interference-canceled signal 
$\vec{z}\R(t)\defn \vec{y}\R(t)  -  \sqrt{\eta\R}\hvec{H}\RR\vec{x}\R(t)$ 
can then be written as 
\begin{align}
  \vec{z}\R(t)
  &= \sqrt{\rho\R}\hvec{H}\sR\vec{x}\s(t)  + \vec{v}\R(t) .	\label{eq:z}
\end{align}
Equation \eqref{z} shows that, in effect, the information signal $\vec{x}\s(t)$ 
propagates through a known channel $\sqrt{\rho\R}\hvec{H}\sR$ corrupted by an 
aggregate (possibly non-Gaussian) noise $\vec{v}\R(t)$, whose 
$(\hvec{H}\sR,\hvec{H}\RR)$-conditional covariance we denote as
$\hvec{\Sigma}\R[l]\defn \cov\{\vec{v}\R(t)\giv\hvec{H}\sR,\hvec{H}\RR\}_{t\in\mc{T}\data[l]}$, recalling that $l\in\{1,2\}$ indexes the data-period.
In \appref{cancellation}, we show that 
\begin{align}
   \hvec{\Sigma}\R[l]
	&\approx  \vec{I} + 
	      \kappa\rho\R\hvec{H}\sR\diag(\vec{Q}\s[l])\hvec{H}\sR\herm
	  +  \hvec{D}\sR\tr(\vec{Q}\s[l]) 
\nonumber\\&\quad
	+  \kappa\eta\R\hvec{H}\RR\diag(\vec{Q}\R[l])\hvec{H}\RR\herm
	  +  \hvec{D}\RR\tr(\vec{Q}\R[l])  		\label{eq:sigma}
\nonumber\\&\quad
	+  \beta\rho\R\diag( \hvec{H}\sR\vec{Q}\s[l]\hvec{H}\sR\herm )
\nonumber\\&\quad
	  +  \beta\eta\R\diag( \hvec{H}\RR\vec{Q}\R[l]\hvec{H}\RR\herm ) ,
\end{align}
where $\hvec{D}\sR\defn \E\{\vec{D}\sR\giv\hvec{H}\sR\}$ 
and $\hvec{D}\RR\defn \E\{\vec{D}\RR\giv\hvec{H}\RR\}$ obey
\begin{align}
   \hvec{D}
	&\approx \frac{1}{2T}
	\bigg(  \vec{I}  +  \alpha \frac{2\kappa}{N} \hvec{H}\hvec{H}\herm
	     + \alpha \frac{2\beta}{N} \diag\Big( \hvec{H} \hvec{H}\herm\Big) 
	\bigg) 	\label{eq:Dhat}
\end{align}
and where the approximations in \eqref{sigma}-\eqref{Dhat} follow from 
$\kappa\ll 1$ and $\beta\ll 1$.
We note, for later use, that the channel estimation error terms 
$\hvec{D}$ can be made arbitrarily small through appropriate 
choice of $T$.

The effective channel from the relay to the destination can be similarly 
stated as 
\begin{align}
  \vec{y}\D(t)
  &= \sqrt{\rho\D}\hvec{H}\RD\vec{x}\R(t) + \vec{v}\D(t)  	\label{eq:yD}\\
  \vec{v}\D(t)
  &\defn   \sqrt{\rho\D}\vec{H}\RD\vec{c}\R(t)
	   -  \vec{D}^{\frac{1}{2}}\RD\tvec{H}\RD \vec{x}\R(t) 
		+  \vec{n}\D(t)  +   \vec{e}\D(t)  
\nonumber\\&\quad
	+ \sqrt{\eta}\D \hvec{H}\sD \big( \vec{x}\s(t) + \vec{c}\s(t)\big) 
	   -  \vec{D}^{\frac{1}{2}}\sD\tvec{H}\sD \big(\vec{x}\s(t) 
\nonumber\\&\quad
	   + \vec{c}\s(t)\big) 
				\label{eq:vD},
\end{align}
and an expression similar to \eqref{sigma} can be derived for 
the destination's aggregate noise covariance,
$\hvec{\Sigma}\D[l] \defn \cov\{\vec{v}\D(t)\giv\hvec{H}\RD,\hvec{H}\sD\}_{t\in\mc{T}\data[l]}$ during data-period $l\in\{1,2\}$.
Unlike the relay node, however, the destination node does not cancel the interference term $\sqrt{\eta}\D \hvec{H}\sD \vec{x}\s(t)$, but rather lumps it in with the aggregate noise $\vec{v}\D(t)$. 
The latter practice is well motivated under the assumption that $\eta\D\ll\rho\R$, i.e., that the source-to-destination link is much weaker than the relay-to-destination link.
\Figref{relay_equiv} summarizes the equivalent system model.

\putFrag{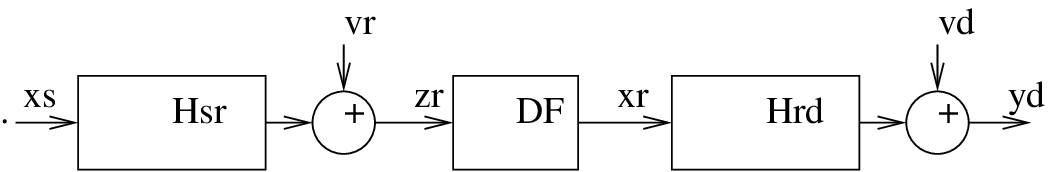}
	{Equivalent model of full-duplex MIMO relaying.}
	{2.8}
        {\newcommand{\sz}{1.0}
         \newcommand{\szz}{0.9}
	 \psfrag{.}{}
	 \psfrag{+}[][Bl][\sz]{$+$}
         \psfrag{Hsr}[][Bl][\szz]{$\sqrt{\rho\R}\hvec{H}\sR$}
	 \psfrag{Hrd}[][Bl][\szz]{$\sqrt{\rho\D}\hvec{H}\RD$}
	 \psfrag{xs}[b][Bl][\sz]{$\vec{x}\s$}
	 \psfrag{vr}[b][Bl][\sz]{$\vec{v}\R$}
	 \psfrag{zr}[b][Bl][\sz]{$\vec{z}\R$}
	 \psfrag{DF}[][Bl][0.48]{\begin{tabular}{c}\textsf{decode \&}\\\textsf{forward}\end{tabular}}
	 \psfrag{xr}[b][Bl][\sz]{$\vec{x}\R$}
	 \psfrag{vd}[b][Bl][\sz]{$\vec{v}\D$}
	 \psfrag{yd}[b][Bl][\sz]{$\vec{y}\D$}
	 }

\subsection{ Bounds on Achievable Rate }	\label{sec:bounds}
The end-to-end mutual information can be written, for a given time-sharing parameter $\tau$, as \cite{Wang:TIT:05}
\begin{align}
 I_{\tau}(\QQ) &= \min\left\{\sum_{l=1}^2 \tau[l] I\sR(\QQ[l]),
		\sum_{l=1}^2 \tau[l] I\RD(\QQ[l])\right\},	
						\label{eq:I}
\end{align}
where $I\sR(\QQ[l])$ and $I\RD(\QQ[l])$ are the period-$l$
mutual informations of the source-to-relay channel 
and relay-to-destination channel, respectively, and where 
$\QQ[l]\defn \big( \vec{Q}\s[l], \vec{Q}\R[l]\big)$
and $\QQ\defn \big( \QQ[1], \QQ[2]\big)$. 

To analyze $I\sR(\QQ[l])$ and $I\RD(\QQ[l])$, we leverage the
equivalent system model shown in \figref{relay_equiv}, 
which includes channel-estimation error and relay-self-interference 
cancellation, and treats the source-to-destination link as a source of noise.
The mutual-information analysis is, however, still complicated by the fact that
the aggregate noises $\vec{v}\R(t)$ and $\vec{v}\D(t)$ are generally 
non-Gaussian, as a result of the channel-estimation-error components 
in \eqref{v} and \eqref{vD}. 
However, it is known that, among all noise distributions of a given
covariance, the Gaussian one is worst 
from a mutual-information perspective \cite{Hassibi:TIT:03}. 
In particular, treating the noise as Gaussian 
yields the lower bounds 
$I\sR (\QQ[l]) \geq \underline{I}\sR(\QQ[l])$
and $I\RD (\QQ[l]) \geq \underline{I}\RD(\QQ[l])$, where \cite{Tse:Book:05}
\begin{align}
  \lefteqn{\underline{I}\sR(\QQ[l])}\nonumber\\
  &=  \log\det\Big( \vec{I} + \rho\R \hvec{H}\sR\vec{Q}\s[l]\hvec{H}\herm\sR \hvec{\Sigma}^{-1}\R[l] \Big) 
		\label{eq:IsR}\\ 
  &= \log\det\Big( \rho\R \hvec{H}\sR\vec{Q}\s[l]\hvec{H}\herm\sR 
		+ \hvec{\Sigma}\R[l] \Big) 
		-\log\det(\hvec{\Sigma}\R[l])    
\end{align}
and
\begin{align}
  \lefteqn{ \underline{I}\RD(\QQ[l]) }\nonumber\\
  &=  \log\det\Big( \vec{I} + \rho\D \hvec{H}\RD\vec{Q}\R[l]\hvec{H}\herm\RD \hvec{\Sigma}^{-1}\R[l] \Big) 
		\label{eq:IRD} \\
  &= \log\det\Big( \rho\D \hvec{H}\RD\vec{Q}\R[l]\hvec{H}\herm\RD 
		+ \hvec{\Sigma}\D[l] \Big) 
		-\log\det(\hvec{\Sigma}\D[l])  , 
\end{align}
and thus a lower bound on the end-to-end $\tau$-specific achievable-rate is
\begin{align}
 \underline{I}_{\tau}(\QQ) 
 &= \min\Bigg\{
 	\underbrace{ \sum_{l=1}^2 \tau[l] \underline{I}\sR(\QQ[l]) 
		}_{\displaystyle \defn \underline{I}\sRt(\QQ)},\,
	\underbrace{ \sum_{l=1}^2 \tau[l] \underline{I}\RD(\QQ[l]) 
		}_{\displaystyle \defn \underline{I}\RDt(\QQ)}
		\Bigg\}.	
						\label{eq:mutinfo}
\end{align}
Moreover, the rate 
$\underline{I}_{\tau}(\QQ)$ bits\footnote{
	Throughout the paper, we take ``$\log$'' to be base-2.}%
-per-channel-use (bpcu) can be achieved 
via independent Gaussian codebooks at the transmitters 
and maximum-likelihood detection at the receivers \cite{Tse:Book:05}.

A straightforward achievable-rate upper bound $\overline{I}_{\tau}(\QQ)$ 
results from the case of perfect CSI (i.e., $\hvec{D}=\vec{0}$),
where $\vec{v}\R(t)$ and $\vec{v}\D(t)$ are Gaussian.
Moreover, the lower bound $\underline{I}_{\tau}(\QQ)$ converges to the upper
bound $\overline{I}_{\tau}(\QQ)$ as the training $T\rightarrow \infty$.

\section{ Transmit Covariance Optimization } 	\label{sec:approach}
We would now like to find the transmit covariance matrices 
$\QQ$ 
that maximize the achievable-rate lower bound $\underline{I}_{\tau}(\QQ)$ 
in \eqref{mutinfo} subject to the per-link power constraint
$\QQ\in\Qcont$, where
\begin{align}
  \Qcont
  \defn \bigg\{
   &\QQ \text{~s.t.~}
   \sum_{l=1}^2
	\!\tau[l] \tr\big(\vec{Q}\s[l]\big) \!\leq\! 1,\, 
   \sum_{l=1}^2
	\!\tau[l] \tr\big(\vec{Q}\R[l]\big) \!\leq\! 1,  
  \nonumber\\&
   \vec{Q}\s[l] = \vec{Q}\s\herm[l] \geq 0,\,
   \vec{Q}\R[l] = \vec{Q}\R\herm[l] \geq 0
  	\bigg\} ,
	\label{eq:constraint}
\end{align}
and subsequently optimize the time-sharing parameter $\tau$.
We note that optimizing the transmit covariance matrices is equivalent to
jointly optimizing the transmission beam-patterns and power levels.
In the sequel, we denote the optimal (i.e., maximin) rate, for a given $\tau$, by
\begin{equation}
 \Ioptt
 \defn \max_{\vec{\QQ}\in\Qcont} \min\big\{ \underline{I}\sRt(\QQ), 
 	\underline{I}\RDt(\QQ) \big\},	
						\label{eq:opt}
\end{equation}
and we use $\Qoptt$ to denote the corresponding set of maximin 
covariance designs $\QQ$ (which are, in general, not unique).
Then, with $\tau_*\defn \arg\max_{\tau\in[0,1]} \Ioptt$, 
the optimal rate is
$\Iopt \defn \underline{I}_{*,\tau_*}$, and the corresponding set of maximin
designs is $\Qopt\defn\mathbb{Q}_{*,\tau_*}$.

\subsection{Weighted-Sum-Rate Optimization}
It is important to realize that, within the maximin design set $\Qoptt$, 
there exists at least one ``link-equalizing'' design, i.e.,
$
  \exists \QQ\in\Qoptt ~~\text{s.t.}~~ 
 	\underline{I}\sRt(\QQ) = \underline{I}\RDt(\QQ).
$
To see why this is the case, notice that, 
given any maximin design $\QQ$ such that 
$\underline{I}\sRt(\QQ) > \underline{I}\RDt(\QQ)$,
a simple scaling of $\vec{Q}\s[l]$ can yield
$\underline{I}\sRt(\QQ) = \underline{I}\RDt(\QQ)$,
and thus an equalizing design.
A similar argument can be made when 
$\underline{I}\RDt(\QQ) > \underline{I}\sRt(\QQ)$.

Referring to the set of \emph{all} link-equalizing designs (maximin or 
otherwise), for a given $\tau$, as
\begin{equation}
  \Qeqt
  \defn \left\{ \QQ\in\Qcont ~~\text{s.t.}~~ 
 	\underline{I}\sRt(\QQ) = \underline{I}\RDt(\QQ)
  	\right\} ,
\end{equation}
the maximin equalizing design can be found by solving either
$\arg\max_{\QQ\in\Qeqt} \underline{I}\sRt(\QQ)$ or\linebreak
$\arg\max_{\QQ\in\Qeqt} \underline{I}\RDt(\QQ)$,
where the equivalence is due to the equalizing property.
More generally, 
the maximin equalizing design can be found by solving
\begin{equation}
  \arg\max_{\QQ\in\Qeqt} \underline{I}_{\tau}(\QQ,\zeta)  		
\label{eq:desired}
\end{equation}
with \emph{any} fixed $\zeta\in[0,1]$ and the $\zeta$-weighted sum-rate 
\begin{equation}
  \underline{I}_{\tau}(\QQ,\zeta)
  \defn \zeta \underline{I}\sRt(\QQ)
 	+ (1-\zeta) \underline{I}\RDt(\QQ) .
\end{equation}

To find the maximin equalizing design,
we propose relaxing the constraint on $\QQ$ from
$\Qeqt$ to $\Qcont$, yielding the $\zeta$-weighted-sum-rate optimization problem
\begin{equation}
  \QQoptt(\zeta)=\arg\max_{\QQ\in\Qcont} \underline{I}_{\tau}(\QQ,\zeta)  .
  								\label{eq:weighted}
\end{equation}
Now, \emph{if} there exists $\zeq\in[0,1]$ such that the solution 
$\QQoptt(\zeq)$ to \eqref{weighted} is link-equalizing, then, 
because $\Qeqt\subset\Qcont$,
we know that $\QQoptt(\zeq)$ must also solve the problem 
\eqref{desired}, implying that $\QQoptt(\zeq)$ is maximin.
\Figref{weighted}(a) illustrates the case where such a $\zeq$ exists.
It may be, however, that no $\zeta\in[0,1]$ yields a link-equalizing solution 
$\QQoptt(\zeta)$, as illustrated in \figref{weighted}(b).
This case occurs when 
$\underline{I}\sRt(\QQoptt(\zeta))>\underline{I}\RDt(\QQoptt(\zeta))$
for all $\zeta\in[0,1]$, such as when $\rho\R \gg \rho\D$.
In this latter case, the maximin rate reduces to
$\Ioptt = \lim_{\zeta\rightarrow 0} \underline{I}\RDt(\QQoptt(\zeta))$.

\putFrag{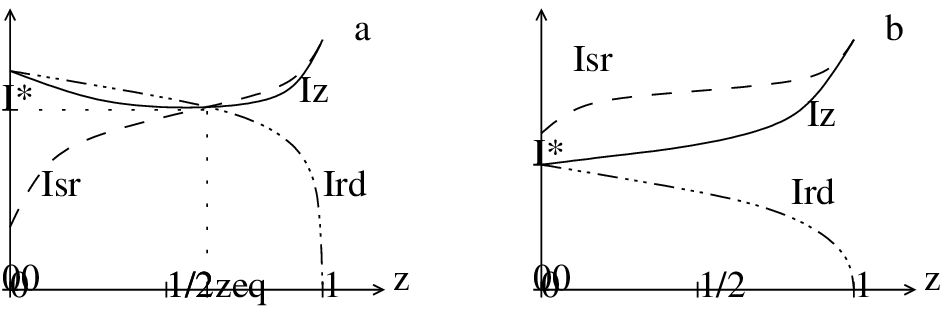}
	{Illustrative examples of $\tau$-specific $\zeta$-weighted sum-rate optimization in the case (a) when a link-equalizing solution exists and (b) when one does not exist.
	 Here, $\underline{I}\sRt(\QQ)$ and $\underline{I}\RDt(\QQ)$ are the source-to-relay and relay-to-destination rates, respectively, $\underline{I}_\tau(\QQ,\zeta)=\zeta\underline{I}\sRt(\QQ)+(1-\zeta)\underline{I}\RDt(\QQ)$ is the $\zeta$-weighted sum-rate, and $\QQoptt(\zeta)$ is the set of optimal covariance matrices for a given time-share $\tau$ and weight $\zeta$.
	 }
	{3.0}
	{\newcommand{\sz}{0.8}
	 \newcommand{\szz}{0.75}
	 \psfrag{a}[Bl][Bl][\sz]{(a)} 
	 \psfrag{b}[Bl][Bl][\sz]{(b)} 
	 \psfrag{00}[r][Bl][\sz]{$0$} 
	 \psfrag{0}[lt][Bl][\sz]{$0$} 
	 \psfrag{1/2}[lt][Bl][\sz]{$\frac{1}{2}$} 
	 \psfrag{1}[lt][Bl][\sz]{$1$} 
	 \psfrag{z}[l][Bl][\sz]{$\zeta$} 
	 \psfrag{zeq}[lt][Bl][\sz]{$\zeta_=$} 
	 \psfrag{I*}[r][Bl][\sz]{$\Ioptt$} 
	 \psfrag{Iz}[l][Bl][\szz]{$\underline{I}_{\tau}(\QQoptt(\zeta),\zeta)$} 
	 \psfrag{Ird}[l][Bl][\szz]{$\underline{I}\RDt(\QQoptt(\zeta))$} 
	 \psfrag{Isr}[l][Bl][\szz]{$\underline{I}\sRt(\QQoptt(\zeta))$} 
	 }

Whether or not $\zeq\in[0,1]$ actually exists, we propose to search for 
$\zeq$ using bisection, leveraging
the fact that $\underline{I}\RDt(\QQoptt(\zeta))$ is non-increasing in $\zeta$
and $\underline{I}\sRt(\QQoptt(\zeta))$ is non-decreasing in $\zeta$.
To perform the bisection search, we initialize the search interval $\mc{I}$ at $[0,1]$, 
and bisect it at each step after testing the condition
$\underline{I}\RDt(\QQoptt(\zeta)) > \underline{I}\sRt(\QQoptt(\zeta))$
at the midpoint location $\zeta$ in $\mc{I}$;
if the condition holds true, we discard the left sub-interval of $\mc{I}$, 
else we discard the right sub-interval.
We stop bisecting when 
$|\underline{I}\RDt(\QQoptt(\zeta)) - \underline{I}\sRt(\QQoptt(\zeta))|$
falls below a threshold or a maximum number of iterations has elapsed.
Notice that, even when there exists no $\zeq\in[0,1]$,
bisection converges towards the desired weight $\zeta=0$.
Subsequently, we optimize over $\tau\in[0,1]$ using a grid-search.

\subsection{Gradient Projection}
At each bisection step, we use Gradient Projection (GP) to solve\footnote{
  Because \eqref{opt} is generally non-convex, finding the global maximum can be difficult.  
  Although GP is guaranteed only to find a local, and not global, maximum,
  our experience with different initializations suggests that GP is indeed 
  finding the global maximum in our problem.}
the $\tau$-specific, $\zeta$-weighted-sum-rate optimization problem \eqref{weighted}.
The GP algorithm \cite{Bertsekas:Book:99} is defined as follows.
For the generic problem of maximizing a function $f(\vec{x})$ over $\vec{x}\in \mathcal{X}$, 
the GP algorithm starts with an initialization $\vec{x}\of{0}$ and iterates the following steps 
for $k=0,1,2,3,\dots$
\begin{align}
  \tvec{x}\of{k} 
  &= \mc{P}_{\mc{X}}\big( \vec{x}\of{k}  +  s\of{k} \nabla f(\vec{x}\of{k}) \big)
		\label{eq:GPalg2}\\
  \vec{x}\of{k+1} 
  &= \vec{x}\of{k} + \gamma\of{k}(\tvec{x}\of{k} - \vec{x}\of{k}) 
		\label{eq:GPalg1} ,
\end{align}
where $\mc{P}_{\mc{X}}(\cdot)$ denotes projection onto the set $\mathcal{X}$ and $\nabla f(\cdot)$
 denotes the gradient of $f(\cdot)$. 
The parameters $\gamma\of{k} \in (0,1]$ and $s\of{k}$ act as stepsizes.
In the sequel, we assume $s\of{k}=1~\forall k$.

In applying GP to the optimization problem \eqref{weighted}, we first take gradient steps for
 $\vec{Q}\R[1]$ and $\vec{Q}\R[2]$, and then project onto the constraint set \eqref{constraint}.
Next, we take gradient steps for $\vec{Q}\s[1]$ and $\vec{Q}\s[2]$, and then project onto the
constraint set.
In summary, denoting the relay gradient by
$\vec{G}\R[l] \defn \nabla_{\vec{Q}\R[l]}\underline{I}_\tau(\QQ,\zeta)$,
our GP algorithm iterates the following steps to convergence:
\begin{align}
  \vec{P}\R\of{k}[1] 
  &= \vec{Q}\R\of{k}[1] + \vec{G}\R\of{k}[1] \label{eq:ourGPbegin} \\
  \vec{P}\R\of{k}[2] 
  &= \vec{Q}\R\of{k}[2] + \vec{G}\R\of{k}[2] \\
  \hspace{-2mm}
  \big(\tvec{Q}\R\of{k}[1],\tvec{Q}\R\of{k}[2]\big)
  &= \mc{P}_{\mc{X}} \big(\vec{P}\R\of{k}[1],\vec{P}\R\of{k}[2]\big) \\
  \vec{Q}\R\of{k+1}[1] 
  &= \vec{Q}\R\of{k}[1] + \gamma\of{k} \big( \tvec{Q}\R\of{k}[1] - \vec{Q}\R\of{k}[1] \big) \\
  \vec{Q}\R\of{k+1}[2] 
  &= \vec{Q}\R\of{k}[2] + \gamma\of{k} \big( \tvec{Q}\R\of{k}[2] - \vec{Q}\R\of{k}[2] \big)  \label{eq:ourGPend}
\end{align}
and then repeats similar steps for $\vec{Q}\s[1]$ and $\vec{Q}\s[2]$.
An outer loop then repeats this pair of inner loops until the maximum change 
in $\QQ$ is below a small positive threshold $\epsilon$.

We now provide additional details on the GP steps.
As for the gradient, \appref{grad_proj} shows that 
the gradient $\vec{G}\R[l]$ can be written as in \eqref{G}, at the top
of the next page,
\begin{figure*}[!t]
 \normalsize
 \begin{IEEEeqnarray}{rCl}
  	\frac{\vec{G}\R[l]}{2\tau[l]}
  	&=& \frac{(1-\zeta)\rho\D}{\ln 2}  \bigg\{	
 	\hvec{H}\herm\RD \Big( \vec{S}\D^{-1}[l] 
	+ \beta \diag\big(\vec{S}\D^{-1}[l] - \hvec{\Sigma}\D^{-1}[l]\big) \Big) 		\hvec{H}\RD
	+ \diag\Big( \kappa \hvec{H}\herm\RD \big( \vec{S}\D^{-1}[l] - \hvec{\Sigma}\D^{-1}[l] \big) \hvec{H}\RD \Big) \bigg\}
  	\nonumber\\&&\quad
	+ \frac{1-\zeta}{\ln 2} \text{sum}\Big(  \hvec{D}\RD^* \odot\big(\vec{S}\D^{-1}[l] 		- \hvec{\Sigma}\D^{-1}[l]\big)\Big) \vec{I}
	+ \frac{\zeta}{\ln 2}\text{sum}\Big(  \hvec{D}\RR^* \odot\big(\vec{S}\R^{-1}[l] - 	\hvec{\Sigma}\R^{-1}[l]\big)\Big) \vec{I}
 	\nonumber\\&&\quad
  	+ \frac{\zeta\eta\R}{\ln 2}  \bigg\{
	\diag \Big(  \kappa \hvec{H}\herm\RR \big(\vec{S}\R^{-1}[l] - \hvec{\Sigma}\R^{-1}[l]\big) \hvec{H}\RR \Big)
	+ \beta \hvec{H}\herm\RD \diag \big(\vec{S}\R^{-1}[l] - \hvec{\Sigma}\R^{-1}[l]\big) \hvec{H}\RD \bigg\} ,
    \IEEElabel{eq:G}
 \end{IEEEeqnarray}
 \hrulefill
 \vspace*{4pt}
\end{figure*}
where 
\begin{align}
\vec{S}\D[l] 
&\defn \rho\D \hvec{H}\RD \vec{Q}\R[l] \hvec{H}\RD\herm + \hvec{\Sigma}\D[l] \label{eq:Sd} \\
\vec{S}\R[l] 
&\defn \rho\R \hvec{H}\sR \vec{Q}\s[l] \hvec{H}\sR\herm + \hvec{\Sigma}\R[l] \label{eq:Sr} .
\end{align}
For $\vec{G}\s[l]$, a similar expression can be derived.

To compute the projection $\mc{P}_{\mc{X}}(\vec{P}\R[1],\vec{P}\R[2])$, we 
first notice that, due to the Hermitian property of $\vec{P}\R[l]$, we can construct an eigenvalue
decomposition $\vec{P}\R[l] = \vec{U}\R[l] \vec{\Lambda}\R[l] \vec{U}\R\herm[l]$ with unitary
$\vec{U}\R[l]$ and real-valued $\vec{\Lambda}\R[l] = \Diag(\lambda_{\textsf{r},1}[l],
\lambda_{\textsf{r},2}[l], \ldots, \lambda_{\textsf{r},N}[l])$.
The projection of $(\vec{P}\R[1],\vec{P}\R[2])$ onto the constraint set \eqref{constraint} then equals
$\tvec{Q}\R[l] = \vec{U}\R[l] (\vec{\Lambda}\R[l] - \mu \vec{I})^{+} \vec{U}\R\herm[l]$, 
where $( \vec{B} )^{+} = \max(\vec{B},\vec{0})$ elementwise, and where $\mu$ is chosen such that
$\sum_{n=1}^{N} \sum_{l=1}^{2} \tau[l] \max( \lambda_{\textsf{r},n}[l] - \mu, 0 ) = 1$.
In essence, $\mc{P}_{\mc{X}}(\cdot)$ performs water-filling.

To adjust the stepsize $\gamma\of{k}$, we use the Armijo stepsize rule \cite{Bertsekas:Book:99}, i.e., 
$\gamma\of{k} = \nu^{m_k}$ where $m_k$ is the smallest nonnegative integer that satisfies
\begin{align}
\lefteqn{ 
	\underline{I}_{\tau}( \QQ\of{k+1},\zeta) - \underline{I}_{\tau}(\QQ\of{k},\zeta) 
} \nonumber\\
	&\geq  \sigma \nu^{m_k} 
			\sum_{l = 1}^{2}
			\tr \!\bigg(  
			\vec{G}\s\ofH{k}[l] \Big(  \tvec{Q}\s\of{k}[l] - \vec{Q}\s\of{k}[l] \Big)
\nonumber\\&\quad
			+ \vec{G}\R\ofH{k}[l] \Big(  \tvec{Q}\R\of{k}[l] - \vec{Q}\R\of{k}[l] \Big)
			\bigg)
\end{align}
for some constants $\sigma,\nu$ typically chosen so that 
$\sigma\in[10^{-5},10^{-1}]$ and $\nu\in[0.1,0.5]$.
Above, we used the shorthand 
$\QQ\of{k}\defn(\vec{Q}\s\of{k}[1], \vec{Q}\s\of{k}[2], \vec{Q}\R\of{k}[1], \vec{Q}\R\of{k}[2])$.

\section{ Achievable-Rate Approximation } 	\label{sec:approx}

The complicated nature of the optimization problem \eqref{opt} motivates
us to approximate its solution, i.e., the covariance-optimized achievable rate 
$\underline{I}_*= \max_{\tau\in[0,1]}\max_{\QQ\in\Qcont}\underline{I}_\tau(\QQ)$. 
In doing so, we focus on the case of $T\rightarrow\infty$, where channel 
estimation error is driven to zero so that
$\underline{I}_\tau(\QQ)=I_\tau(\QQ)=\overline{I}_\tau(\QQ)$.
In addition, for tractability, we restrict ourselves to the case $\Ns=\NR=N$ and 
$\MR=\MD=M$ (i.e., $N$ transmit antennas and $M$ receive antennas at each node),
the case $\eta\D=0$ (i.e., no direct source-to-destination link), and
the case $\tau=\frac{1}{2}$ (i.e., equal time-sharing).

Our approximation is built around the simplifying case that the channel matrices 
$\{\vec{H}\sR,\vec{H}\RR,\vec{H}\RD\}$ are each diagonal, 
although not necessarily square, and have $R\defn\min\{M,N\}$ identical 
diagonal entries equal to $\sqrt{MN/R}$.
(The latter value is chosen so that 
$\E\{\tr(\vec{H}\vec{H}\herm)\}=MN$ 
as assumed in \secref{chan}.)
In this case, the mutual information \eqref{mutinfo} becomes \eqref{approx1}, 
at the top of the next page.
\begin{figure*}[!t]
 \normalsize
 \begin{IEEEeqnarray}{rCl}
 	 I_{\tau}(\QQ)
  &\approx& \frac{1}{2}\min\bigg\{
  	\sum_{l=1}^2 \log\det \bigg( \vec{I} + \rho\R 
  	\frac{NM}{R} 
	\vec{Q}\s[l] \Big( \vec{I} + (\kappa+\beta) 
	\frac{NM}{R} 
	\big( \rho\R \diag(\vec{Q}\s[l])
	+ \eta\R \diag(\vec{Q}\R[l])
  	\big) \Big)^{-1} \bigg) ,
\nonumber\\&&\quad
  	\sum_{l=1}^2 \log\det \bigg( \vec{I} + \rho\D 
  	\frac{NM}{R} 
	\vec{Q}\R[l] \Big( \vec{I} + (\kappa+\beta) 
	\frac{NM}{R} 
	\rho\D \diag(\vec{Q}\R[l])
	\Big)^{-1} \bigg) 
	\bigg\} . 
    \IEEElabel{eq:approx1}
 \end{IEEEeqnarray}
 \hrulefill
 \vspace*{4pt}
\end{figure*} 
When $\eta\R\ll\rho\R$,
the $\eta\R$-dependent 
terms in \eqref{approx1}
can be ignored, after which it is straightforward to show that, under
the constraint \eqref{constraint}, the optimal
covariances are the ``full duplex'' 
$\QQ\fd\defn(\frac{1}{N}\vec{I},\frac{1}{N}\vec{I},
\frac{1}{N}\vec{I},\frac{1}{N}\vec{I})$, 
for which \eqref{approx1} gives
\begin{align}
  \lefteqn{I(\QQ\fd)}\nonumber\\
  &\approx\, R \log \left( 1 + 
  	\min\Big\{  \textstyle
		\frac{\rho\R}{\frac{R}{M} + (\kappa+\beta)(\rho\R+\eta\R)} ,
  		\frac{\rho\D}{\frac{R}{M} + (\kappa+\beta)\rho\D} 
	\Big\} \right) \\
  &= \begin{cases}
  	R \log \Big( 1 + \frac{\rho\D}{\frac{R}{M} 
  	+ (\kappa+\beta)\rho\D}\Big) 
	& \text{if~} 
	\frac{\rho\R}{\rho\D} \geq 1\!+\! \frac{(\kappa+\beta) \eta\R M}{R} \\
  	R \log \Big( 1 + \frac{\rho\R}{\frac{R}{M}
  	+ (\kappa+\beta)(\rho\R+\eta\R)}\Big) 
	& \text{else}.
    \end{cases} 	\label{eq:Ifd} 
\end{align}
When $\eta\R\gg\rho\R$, the $\eta\R$-dependent term in 
\eqref{approx1} dominates unless $\vec{Q}\R[l]=\vec{0}$.
In this case, the optimal covariances are the ``half duplex'' ones
$\QQ\hd\defn(\frac{2}{N}\vec{I},\vec{0},\vec{0},\frac{2}{N}\vec{I})$, 
for which \eqref{approx1} gives
\begin{align}
  I(\QQ\hd) 
  &\approx \left\{ \begin{array}{@{}l@{~}l@{}}
  	\frac{R}{2} \log \Big( 1 + \frac{\rho\D}{\frac{R}{2M} 
  	+ (\kappa+\beta)\rho\D}\Big) 
	& \text{if~} \frac{\rho\R}{\rho\D}\geq 1 \\
  	\frac{R}{2} \log \Big( 1 + \frac{\rho\R}{\frac{R}{2M}
  	+ (\kappa+\beta)\rho\R}\Big) 
	& \text{else}.
  	\end{array} \right. 
  \label{eq:Ihd} 
\end{align}
Finally, given any triple $(\rho\R,\eta\R,\rho\D)$, we approximate the 
achievable rate as follows: 
$I_*\approx \max\{I(\QQ\fd),I(\QQ\hd)\}$.

From \eqref{Ifd}-\eqref{Ihd}, using $\theta \defn \frac{R}{M(\kappa+\beta)}$,
it is straightforward to show that the approximated system operates as follows.
\begin{enumerate}
\item
Say $\frac{\rho\R}{\rho\D}\leq 1$. Then full-duplex is used iff
\begin{equation}
 \eta\R
 \leq \frac{1}{2}\sqrt{(\theta+2\rho\R)^2+\frac{2\rho\R}{\kappa+\beta}(\theta+2\rho\R)}
 	-\frac{1}{2}\theta .
 \label{eq:bndry1}
\end{equation}
For either half- or full-duplex, $I_*$ is invariant to $\rho\D$, i.e.,
the source-to-relay link is the limiting one.
\item
Say $1\leq \frac{\rho\R}{\rho\D}\leq 1+\frac{(\kappa+\beta)\eta\R M}{R}$.
Full-duplex is used iff
\begin{equation}
 \eta\R
 \leq 
 \frac{\rho\R}{2\rho\D}
 \sqrt{(\theta+2\rho\D)^2+\frac{2\rho\D}{\kappa+\beta}(\theta+2\rho\D)}
 	-\theta\Big(1-\frac{\rho\R}{2\rho\D}\Big) .
 \label{eq:bndry2}
\end{equation}
\item
Say $1+\frac{(\kappa+\beta)\eta\R M}{R}\leq \frac{\rho\R}{\rho\D}$, or equivalently
$\eta\R \leq \eta\crit \defn \big(\frac{\rho\R}{\rho\D}-1\big)\frac{R}{M(\kappa+\beta)}$.
Then full-duplex is always used, and $I_*$ is invariant to $\rho\R$ and $\eta\R$, i.e.,
the rate is limited by the relay-to-destination link.
\end{enumerate}

\Figref{minrate_approx} shows a contour plot of the proposed
achievable-rate approximation as a function of INR $\eta\R$ and 
SNR $\rho\R$, for the case that $\rho\R/\rho\D=2$. 
We shall see in \secref{sims} that our approximation of the 
covariance-optimized achievable-rate is reasonably close 
to that found by solving \eqref{opt} using bisection/GP.

\putFrag{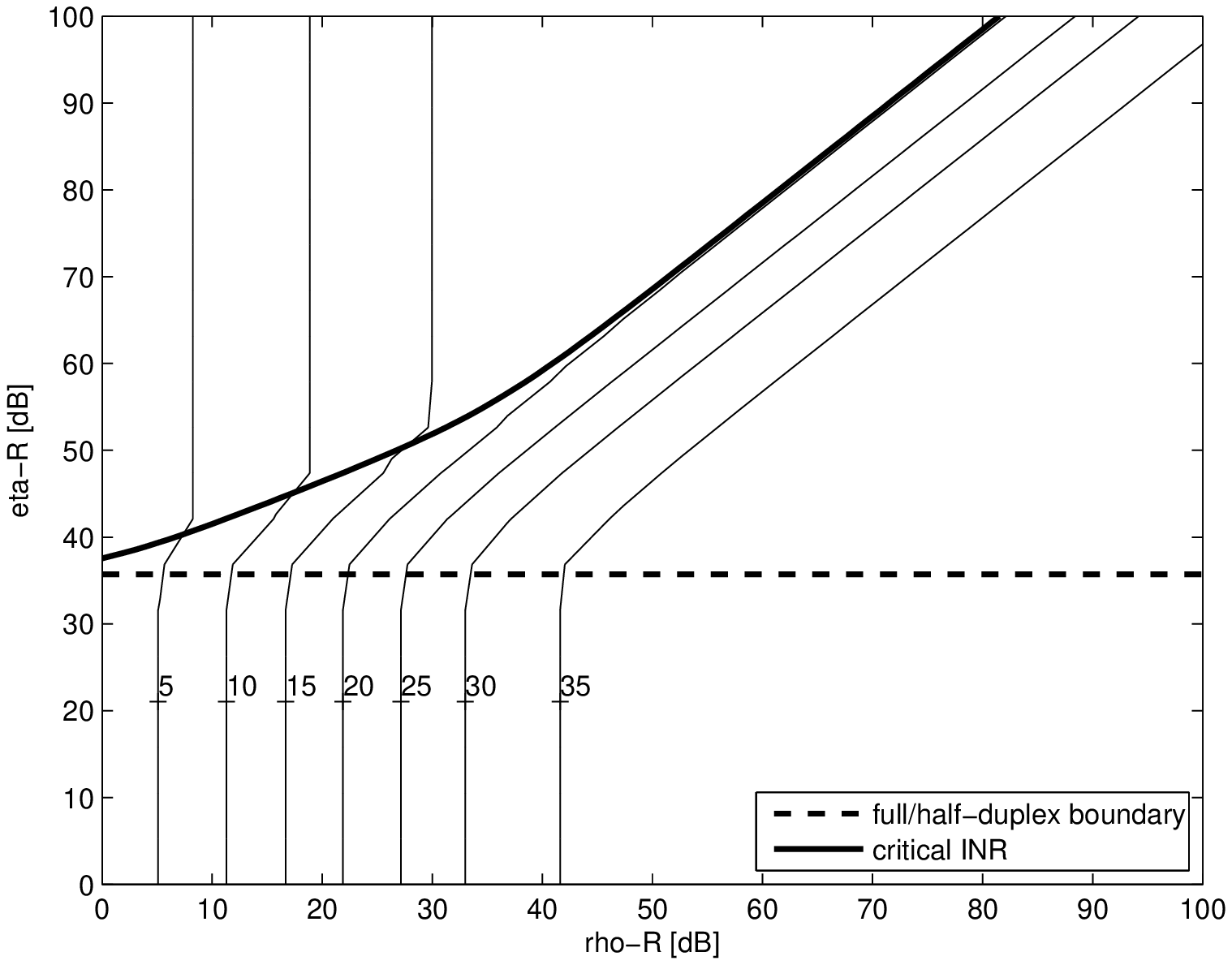}
	{Contour plot of the approximated achievable rate $I_*$
	 versus relay SNR $\rho\R$ and INR $\eta\R$, for $N=3$, $M=4$, 
	 $\beta=\kappa=-40$dB, and $\rho\R/\rho\D = 2$.
	 The horizontal dashed line shows the INR $\eta\crit$,
	 and the dark curve shows the boundary between full- and half-duplex 
	 regimes described in \eqref{bndry2}.
	 }
	{\figsize}
	{\psfrag{eta-R [dB]}[][][0.7]{\sf INR $\eta\R$ [dB]} 
	 \psfrag{rho-R [dB]}[][][0.7]{\sf SNR $\rho\R$ [dB]}}

\section{ Numerical Results }				\label{sec:sims}
In this section, we numerically investigate the behavior of the end-to-end 
rates achievable for full-duplex MIMO relaying under the proposed 
limited transmitter/receiver-DR and channel-estimation-error models.
Recall that, in \secref{analysis}, it was shown that,
for a fixed set of transmit covariance matrices $\QQ$ 
and time-sharing parameter $\tau$,
the achievable rate $I_{\tau}(\QQ)$ can be lower-bounded using 
$\underline{I}_{\tau}(\QQ)$ from \eqref{mutinfo}, and upper-bounded using the
perfect-CSI $\overline{I}_{\tau}(\QQ)$, where the
bounds converge as training $T\rightarrow\infty$. 
Then, in \secref{approach}, a bisection/GP scheme was proposed to 
maximize $\underline{I}_{\tau}(\QQ)$ subject to the power-constraint 
$\QQ\in\Qcont$, which was subsequently maximized over $\tau\in[0,1]$.

We now study the average behavior of the bisection/GP-optimized rate
$\Iopt=\max_\tau\max_{\QQ\in\Qcont}\underline{I}_{\tau}(\QQ)$ as
a function of SNRs $\rho\R$ and $\rho\D$; INRs $\eta\R$ and $\eta\D$; 
dynamic range parameters $\kappa$ and $\beta$; 
number of antennas $N\s$, $N\R$, $M\R$, and $M\D$; and training length $T$.
We also investigate the role of interference cancellation,
the role of two distinct data periods,
the role of $\tau$-optimization,
and the relation to optimized half-duplex (OHD) signaling. 
In doing so, we find close agreement with the achievable-rate
approximation proposed in \secref{approx} and illustrated in 
\figref{minrate_approx}.

For the numerical results below, 
the propagation channel model from 
\secref{chan} and the limited transmitter/receiver-DR models from 
\secref{lim_tdr} and \secref{lim_rdr} were employed, pilot-aided channel 
estimation was implemented as in \secref{chan_est}, and the power 
constraint \eqref{constraint} was applied, implying
the channel-estimation-error covariance \eqref{chan_est_err} and 
the aggregate-noise covariance \eqref{sigma}.
Throughout, we used 
$N \defn N\s = N\R$ transmit antennas,
$M \defn M\R = M\D$ receive antennas,
the SNR ratio $\rho\R/\rho\D = 2$,
the destination INR $\eta\D = 1$, 
training duration $T = 50$ (as justified below),
Armijo parameters $\sigma = 0.01$ and $\nu = 0.2$, 
and GP stopping threshold $\epsilon = 0.01$. 
For each channel realization, the time-sharing coefficient $\tau$ was optimized over the grid $\tau \in \{0.1,0.2,0.3,\dots,0.9\}$, and all results were averaged over $100$ realizations unless specified otherwise.

Below, we denote the full scheme proposed in \secref{approach} by 
``TCO-2-IC,'' which indicates the use of interference cancellation (IC) 
and transmit covariance optimization (TCO) performed 
individually over the 2 data periods
(i.e., $\mc{T}\data[1]$ and $\mc{T}\data[2]$).
To test the impact of IC and of two data periods, 
we also implemented the proposed scheme but without IC, 
which we refer to as ``TCO-2,''
as well as the proposed scheme with only one data period 
(i.e., $\vec{Q}_i[1]=\vec{Q}_i[2]~\forall i$), which we refer 
to as ``TCO-1-IC.'' 
To optimize\footnote{ 
We note that \emph{both} half-duplex and the proposed TCO-2-IC scheme could potentially benefit from allowing the relay to change the partitioning of antennas from transmission to reception across the data period $l\in\{1,2\}$. 
In half duplex mode, for example, it would be advantageous for the relay to use 
$(N\R[1],M\R[1])=(0,7)$ and $(N\R[2],M\R[2])=(7,0)$ as opposed to 
$(N\R[l],M\R[l])=(3,4)~\forall l$.
We do not consider such antenna-swapping in this work, however. 
} half-duplex, we used GP to maximize the sum-rate
$\underline{I}_{\tau}(\QQ,\frac{1}{2})$ under the power constraint 
\eqref{constraint} and
the half-duplex constraint $\vec{Q}_1[2]=\vec{0}=\vec{Q}_2[1]$;
$\tau$-optimization was performed as described above.

To mitigate GP's sensitivity to initialization, we tried two initializations 
for each $\zeta$-weighted-sum-rate problem, OHD and ``naive'' full-duplex (NFD), 
and the one yielding the maximum min-rate was retained.
OHD was calculated as explained above, whereas NFD employed non-zero
OHD covariance matrices $\vec{Q}_1[1]$ and $\vec{Q}_2[2]$ over both data periods
(which is indeed optimal when $\eta\R=0 =\eta\D$). 
Note that both OHD and NFD are invariant to $\zeta$, $\eta\R$, and $\eta\D$.

In \figref{Training_bw}, we investigate the role of channel-estimation training length $T$ 
on the achievable-rate lower bound $\underline{I}(\QQ)$ of TCO-2-IC.
There we see that the rate increases rapidly in $T$ for small values of $T$, but quickly saturates 
for larger values of $T$.
This behavior can be understood from \eqref{sigma}-\eqref{Dhat}, which suggest that channel 
estimation error will have a negligible effect 
on the noise covariances $\hvec{\Sigma}\R[l]$ and $\hvec{\Sigma}\D[l]$ when $T N \gg 1$. 
\Figref{Training_bw} also shows the corresponding achievable-rate upper bounds $\overline{I}(\QQ)$. 
These traces confirm that the nominal training length $T = 50$ ensures 
$\underline{I}(\QQ) \approx \overline{I}(\QQ) \approx I(\QQ)$.

\putFrag{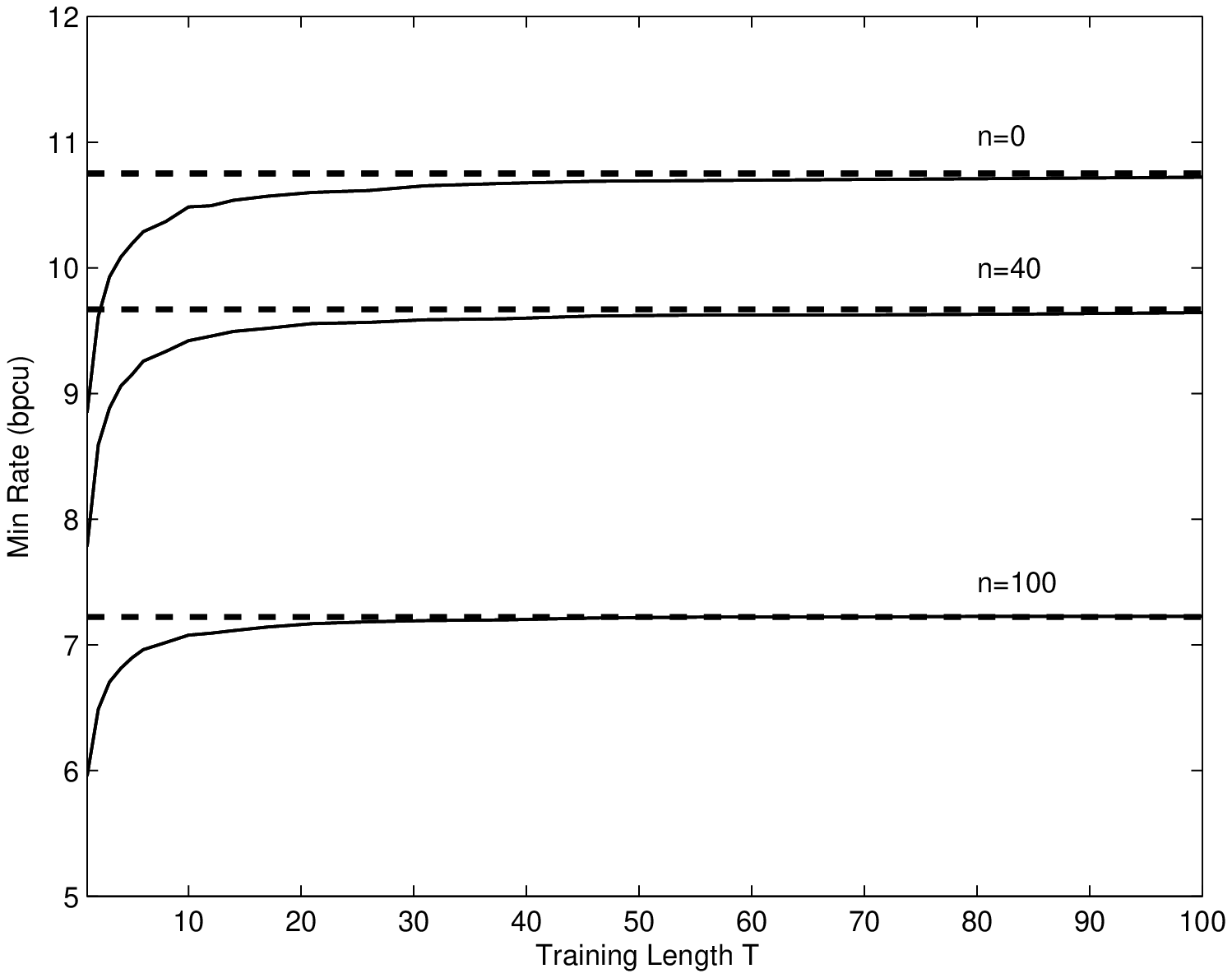}{
Achievable-rate lower bound $\underline{I}_*$ for TCO-2-IC versus training interval $T$.
Here, $N = 3$, $M = 4$, $\beta = \kappa = -40\dB$, $\rho\R = 15\dB$, $\rho\R/\rho\D = 2$, and $\eta\D = 0\dB$.
Also shown as a dashed line which is the corresponding upper bound $\overline{I}_*$ for each value of $\eta\R$.
}{\figsize}{
	\psfrag{n=0}[l][l][0.7]{\sf $\eta\R \!=\! 0\dB$}
	\psfrag{n=40}[l][l][0.7]{\sf $\eta\R \!=\! 40\dB$}
	\psfrag{n=100}[l][l][0.7]{\sf $\eta\R \!=\! 100\dB$}
	\psfrag{Training Length T}[][][0.7]{\sf Training Length $T$}
	\psfrag{Min Rate (bpcu)}[][][0.7]{\sf Min Rate (bpcu)}
}

In \figref{rate_vs_eta_bw}, we examine achievable-rate performance versus INR $\eta\R$ 
for the TCO-2-IC, TCO-1-IC, TCO-2, and OHD schemes, 
using different dynamic range parameters $\beta = \kappa$. 
For OHD, we see that rate is invariant to INR $\eta\R$, as expected.
For the proposed TCO-2-IC, we observe ``full duplex'' performance for 
low-to-mid values of $\eta\R$ and a transition to OHD performance at high 
values of $\eta\R$, just as predicted by the approximation in \secref{approx}.
In fact, the rates in \figref{rate_vs_eta_bw} are very close to 
the approximated values in \figref{minrate_approx}.
To see the importance of two distinct data-communication periods, we examine the 
TCO-1-IC trace, where we observe TCO-2-IC-like performance at low-to-midrange 
values of $\eta\R$, but performance that drops below OHD at high $\eta\R$. 
Essentially, TCO-1-IC forces full-duplex signaling at high INR
$\eta\R$, where half-duplex signaling is optimal, while
TCO-2-IC facilitates the possibility of half-duplex 
signaling through the use of two distinct data-communication periods, 
similar to the MIMO-interference-channel scheme in \cite{Rong:TWC:08}.
The effect of $\tau$-optimization can be seen by comparing the two OHD traces, 
one which uses the fixed value $\tau=0.5$ and the other which uses the optimized 
value $\tau=\tau_*$.
The separation between these traces shows that $\tau$-optimization gives a 
small but noticable rate gain.
Finally, by examining the TCO-2 trace, we conclude that partial interference 
cancellation is very important for all but extremely low or high values of INR $\eta\R$.

\putFrag{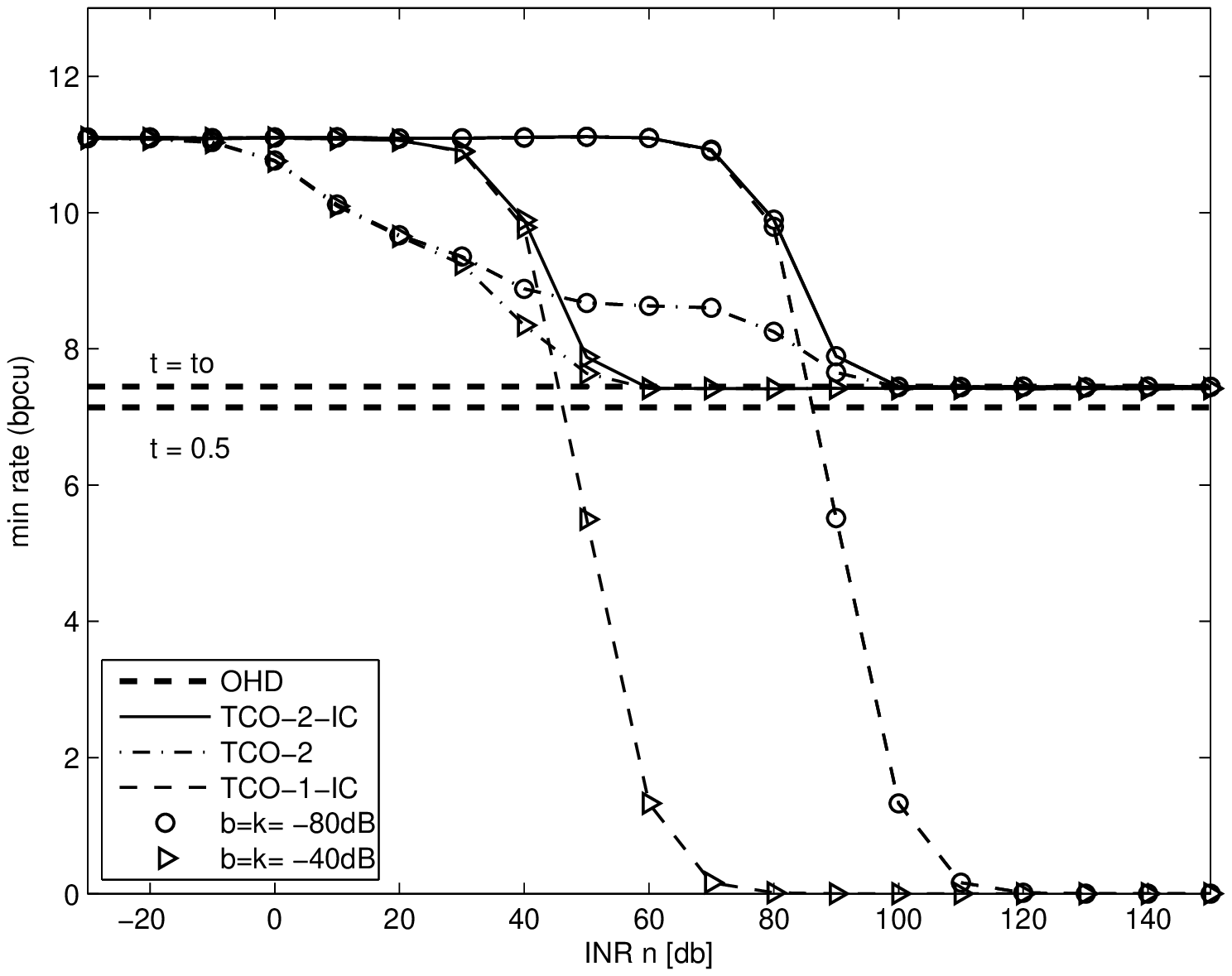}{
Achievable-rate lower bound $\underline{I}_*$ for TCO-2-IC, TCO-2, TCO-1-IC, and OHD versus INR $\eta\R$. Here, $N = 3$, $M = 4$, $\rho\R = 15\dB$, $\rho\R/\rho\D = 2$, $\eta\D = 0\dB$, and $T = 50$. 
OHD is plotted for $\beta = \kappa = -40\dB$, but was observed to give nearly identical rate for $\beta = \kappa=-80$dB. 
Both fixed-time-share ($\tau=0.5$) and optimized-time-share ($\tau=\tau_*$) versions of OHD are shown.}{\figsize}
{
	\psfrag{b=k= -80dB}[l][l][0.54]{\sf $\beta \!=\! \kappa \!=\! -80\dB$} 	
	\psfrag{b=k= -40dB}[l][l][0.54]{\sf $\beta \!=\! \kappa \!=\! -40\dB$}
	\psfrag{t = to}[B][][0.7]{\sf \:$\tau = \tau_\star$}
	\psfrag{t = 0.5}[][t][0.7]{\sf $\tau = 0.5$}
	\psfrag{INR n [db]}[][][0.7]{\sf INR $\eta\R$ (dB)}
	\psfrag{HD}[l][l][0.67]{\sf OHD}
	\psfrag{min rate (bpcu)}[][][0.7]{\sf Min Rate (bpcu)}
}
                      
In \figref{rate_vs_rho_bw},
we examine the rate of the proposed TCO-IC-2 and OHD versus SNR $\rho\R$,
using the dynamic range parameters $\beta=\kappa=-40$dB,
$\eta\D = 0$dB, and two fixed values of INR $\eta\R$.
All the behaviors in \figref{rate_vs_rho_bw} are predicted
by the rate approximation described in \secref{approx} and 
illustrated in \figref{minrate_approx}.
In particular, 
at the low INR of $\eta\R=20$dB, TCO-IC-2 operates in the full-duplex regime
for all values of SNR $\rho\R$.
Meanwhile, at the high INR of $\eta\R=60$dB, TCO-IC-2 operates in half-duplex
at low values of SNR $\rho\R$, but switches to full-duplex after 
$\rho\R$ exceeds a threshold.

\putFrag{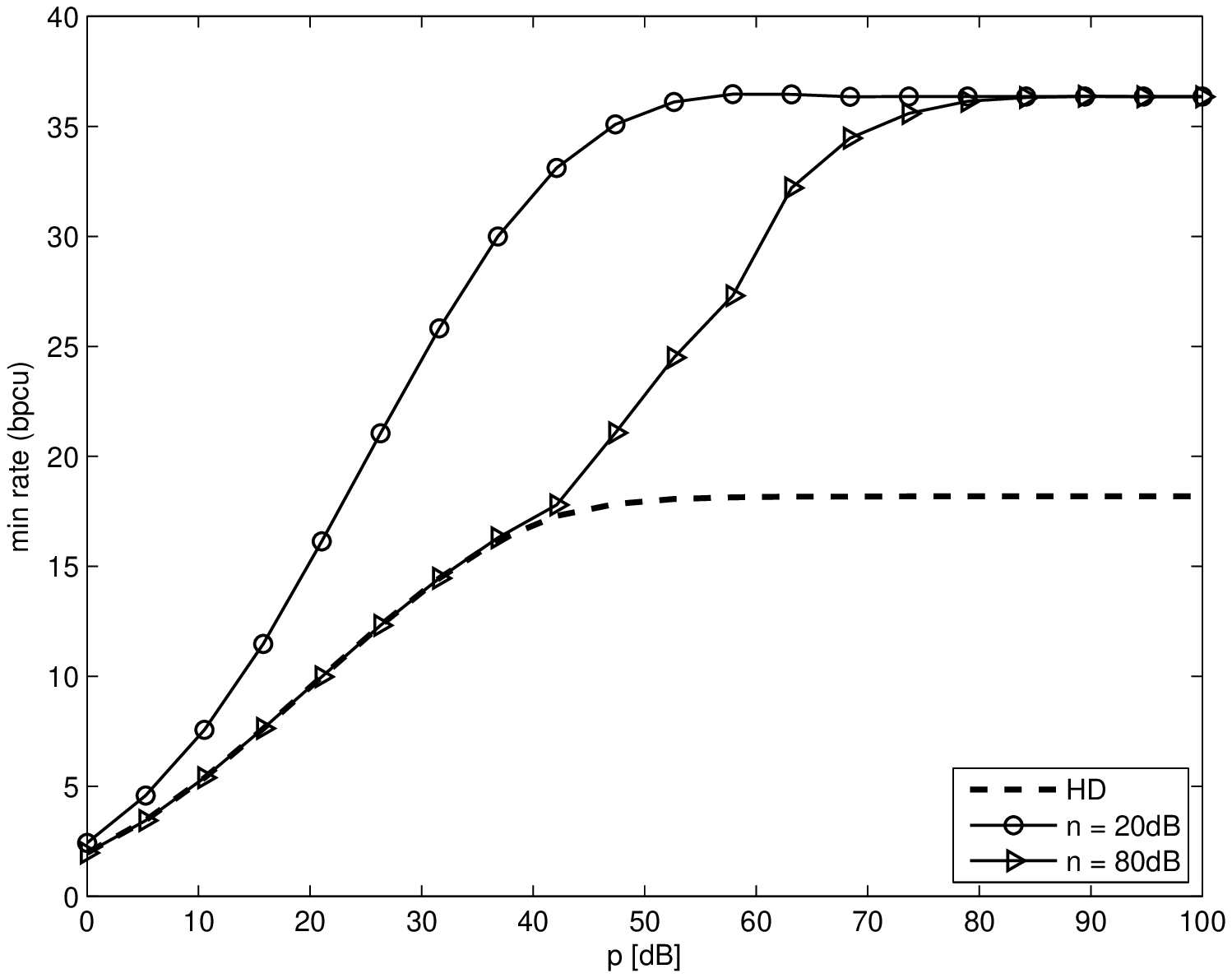}{
Achievable-rate lower bound $\underline{I}_*$ for TCO-2-IC and OHD versus SNR $\rho\R$.
Here, $\rho\R/\rho\D = 2$, $\eta\D = 0\dB$, $N = 3$, $M = 4$, $\beta=\kappa = -40\dB$, and $T = 50$.
OHD in this figure is optimized over $\tau$.}{\figsize}
{
 	\psfrag{n = 20dB}[l][l][0.6]{\sf $\eta\R = 20\dB$}	
 	\psfrag{n = 80dB}[l][l][0.6]{\sf $\eta\R = 80\dB$}	
	\psfrag{p [dB]}[][][0.7]{\sf SNR $\rho\R$ (dB)}	
	\psfrag{HD}[l][l][0.67]{\sf OHD}
	\psfrag{min rate (bpcu)}[][][0.7]{\sf Min Rate (bpcu)}
}

In \figref{minrate}, we plot the 
GP-optimized rate contours of the proposed 
TCO-IC-2 versus both SNR $\rho\R$ and INR $\eta\R$, for comparison to the
approximation in \figref{minrate_approx}.
The two plots show a relatively good match, confirming the accuracy
of the approximation.
The greatest discrepancy between the plots occurs when $\eta\R\approx\rho\R$
and both $\eta\R$ and $\rho\R$ are large, 
which makes sense because the approximation was derived using 
$\eta\R\ll\rho\R$ and $\eta\R\gg\rho\R$.

\putFrag{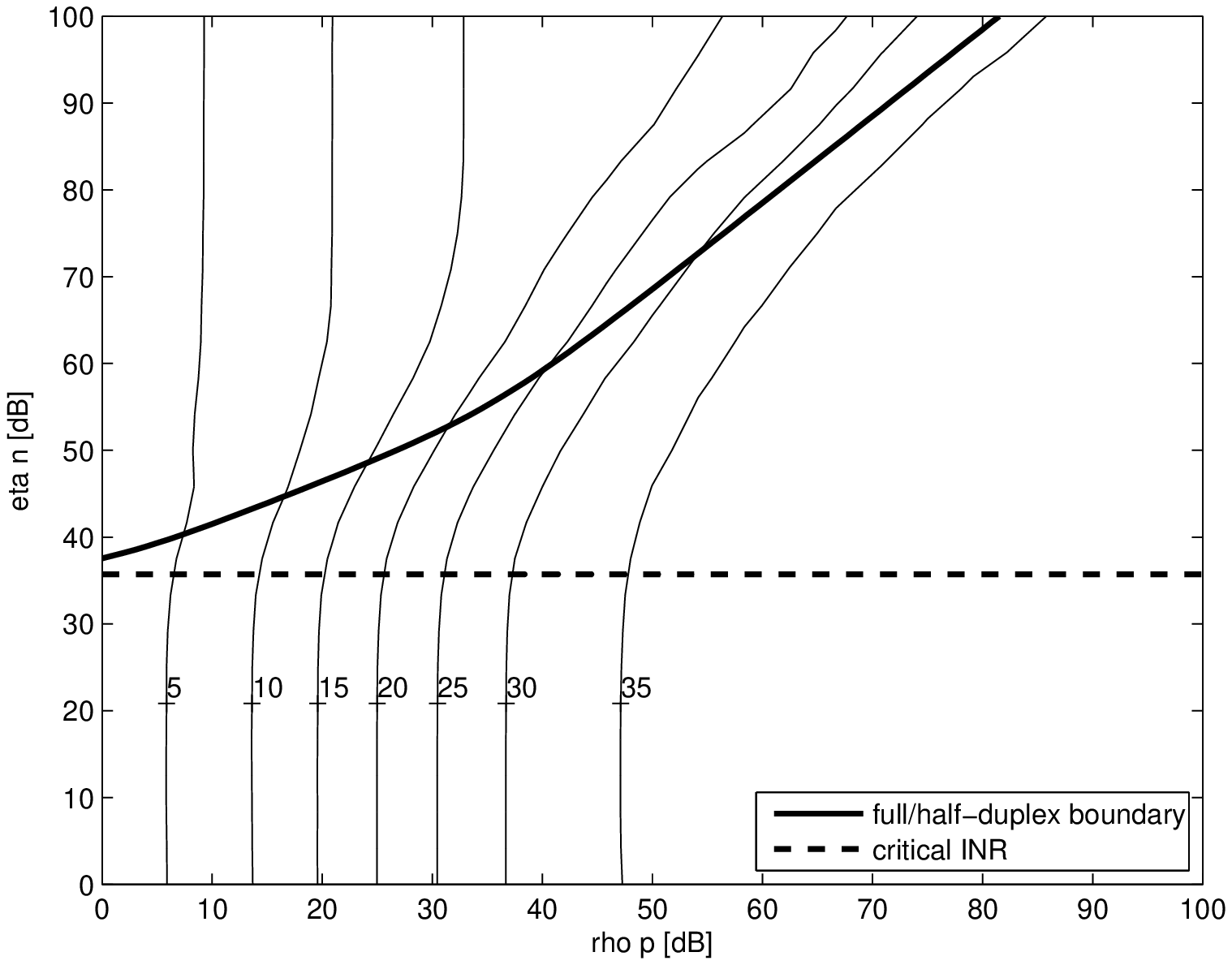}
	{Contour plot of the achievable-rate lower bound $\underline{I}_*$ 
	 for TCO-2-IC versus INR $\eta\R$ and SNR $\rho\R$, for $\rho\D=\rho\R/2$, $\eta\D = 0\dB$,
	 $N=3$, $M=4$, and $\beta=\kappa=-40$dB. 
	 The dark curve (i.e., approximate full/half-duplex boundary) and dashed line 
	 (i.e., critical INR $\eta\crit$) are the same as in \figref{minrate_approx}, and shown for reference. The results
	 are averaged over 250 realizations.}
	{\figsize}
	{\psfrag{eta n [dB]}[][][0.7]{\sf INR $\eta\R$ (dB)} 
	 \psfrag{rho p [dB]}[][][0.7]{\sf SNR $\rho\R$ (dB)}
	 }


Finally, in \figref{Increase_NT_bw}, we explore the achievable rate of 
TCO-2-IC and OHD versus the number of antennas, $N$ and $M$, 
for fixed values of SNR $\rho\R=15\dB$ and $\rho\R/\rho\D=2$, 
INR $\eta\R=30\dB$ and $\eta\D=0\dB$, and 
DR parameters $\beta=\kappa=-40\dB$.
We recall, from \figref{rate_vs_eta_bw}, that these parameters correspond to 
the interesting regime where TCO-2-IC performs between half- and full-duplex.
In \figref{Increase_NT_bw}, 
we see that achievable rate increases with both $M$ and $N$ numbers of antennas, as expected.
More interesting is the achievable-rate behavior when the total number of 
antennas per modem is fixed, e.g., at $N+M=7$, as illustrated
by the triangles in \figref{Increase_NT_bw}.
The figure indicates that the configurations $(N,M)=(3,4)$ and $(N,M)=(4,3)$ 
are best, which (it can be shown) is consistent with approximation 
from \secref{approx}.

\putFrag{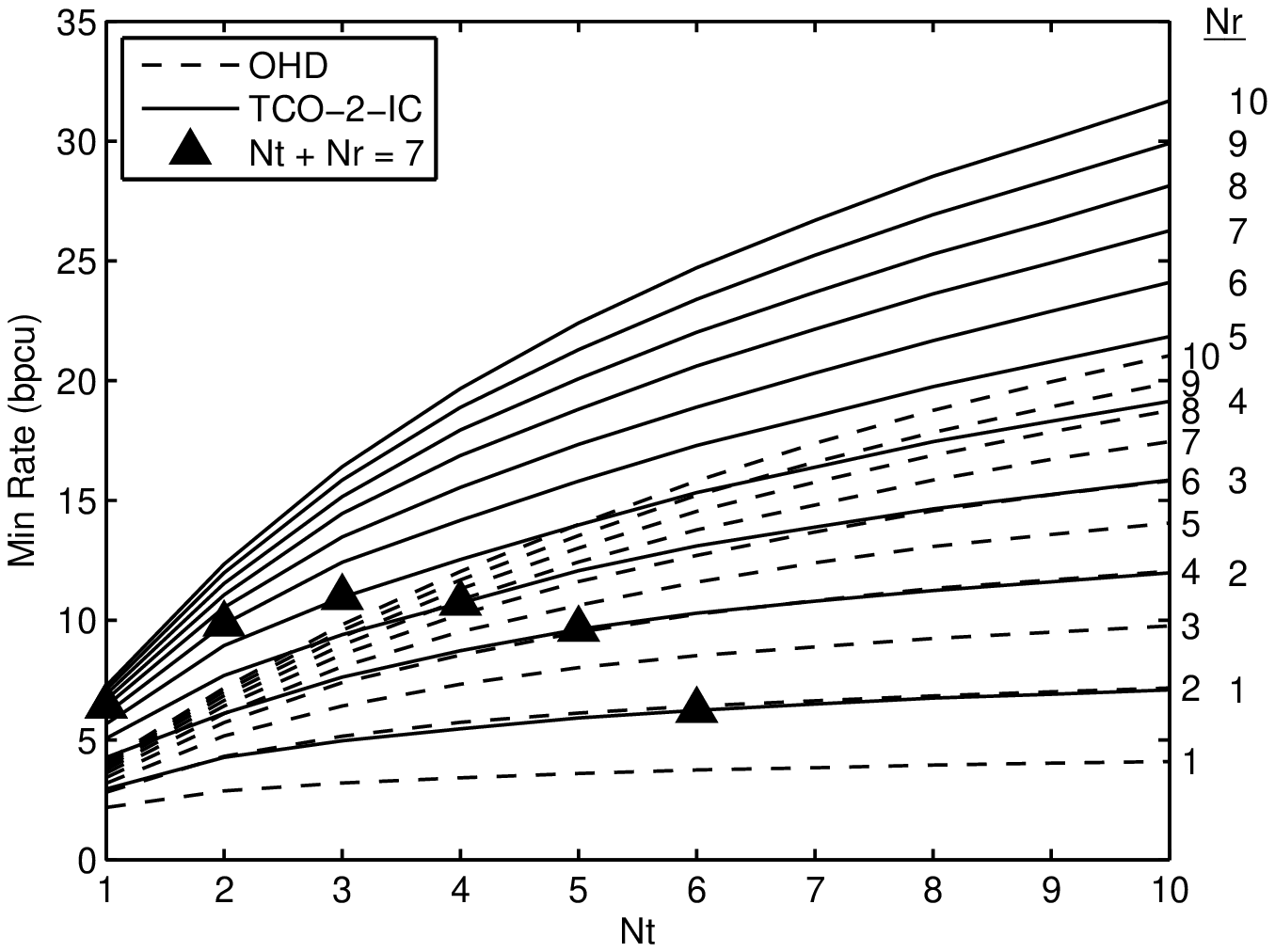}{  Achievable-rate lower bound $\underline{I}_*$ for TCO-2-IC and OHD versus number of transmit antennas $N$ with various numbers of receive antennas $M$.  Here, 
$\rho\R = 15\dB$, $\rho\R/\rho\D=2$, $\eta\R = 30\dB$, $\eta\D = 0\dB$, $\beta = \kappa = -40\dB$, and $T = 50$. 
OHD shown in this figure is optimized over $\tau$.
}{ \figsize }
	{ 	\psfrag{Nt}[][][0.7]{\sf Number of transmit antennas $N$}
		\psfrag{Nr}[Bl][Bl][0.85]{$M$}
		\psfrag{Nt + Nr = 7}[l][l][0.8]{\!$N\!+\!M\!=\!7$}
		\psfrag{HD}[l][l][0.67]{\sf OHD}
 	}

\section{Conclusion}					\label{sec:conclusion}
We considered the problem of decode-and-forward-based full-duplex MIMO relaying 
between a source node and destination node.
In our analysis, we considered limited transmitter/receiver dynamic range, 
imperfect CSI, background AWGN, and very high levels of self-interference.
Using explicit models for dynamic-range limitation and pilot-aided channel 
estimation error, we derived upper and lower bounds on the end-to-end achievable 
rate that tighten as the number of pilots increases.
Furthermore, we proposed a transmission scheme based on maximizing the
achievable-rate lower-bound. 
The latter requires the solution to a nonconvex optimization problem, for which
we use bisection search and Gradient Projection, the latter of which 
implicitly performs water-filling.
In addition, we derived an analytic approximation to the achievable rate 
that agrees closely with the results of the numerical optimization.
Finally, we studied the achievable-rate numerically, as a function of signal-to-noise
ratio, interference-to-noise ratio, transmitter/receiver dynamic range,
number of antennas, and number of pilots.
In future work, we plan to investigate the effect of 
practical coding/decoding schemes, 
channel time-variation, and 
bidirectional relaying.

\appendices
\section{Channel Estimation Details}			\label{app:chan_est}
 In this appendix, we derive certain details of \secref{chan_est}.
 Under limited transmitter-DR, the undistorted received space-time signal is
\begin{equation}
   \vec{U}
   = \sqrt{\alpha}\vec{H}(\vec{X}+\vec{C}) + \vec{N} ,
\end{equation}
 where the spatial correlation\footnote{
   The spatial correlation of $\vec{X}=[\vec{x}(1),\dots,\vec{x}(TN)]$ 
   is $\E\{\vec{x}(t)\vec{x}(t)\herm\}
   =\E\{\frac{1}{TN}\sum_{t=1}^{TN} \vec{x}(t)\vec{x}(t)\herm\}
    = \E\{\frac{1}{T N}\vec{XX}\herm\}$.}
 of the non-distorted pilot 
 signal $\vec{X}$ equals $\frac{2}{N}\vec{I}$ and hence the spatial 
 correlation of the transmitter distortion $\vec{C}$ equals 
 $\frac{2\kappa}{N}\vec{I}$.
 Conditioned on $\vec{H}$, the spatial correlation of $\vec{U}$ is then
 $ \vec{\Phi}
   = \frac{2\alpha(1+\kappa)}{N}\vec{HH}\herm + \vec{I} $,
 and hence the $\vec{H}$-conditional spatial correlation of the 
 receiver distortion $\vec{E}$ equals
\begin{equation}
   \beta\diag(\vec{\Phi}) 
   = \beta\bigg(
   	\frac{2\alpha(1+\kappa)}{N}\diag\Big(\vec{HH}\herm\Big) + \vec{I}\bigg).
\end{equation}
 Given \eqref{Y}, the distorted received signal $\vec{Y}$ can be
 written as
\begin{equation}
   \vec{Y}
   = \sqrt{\alpha}\vec{H}\vec{X} + \vec{W} ,
\end{equation}
 where $\vec{W} \defn \sqrt{\alpha}\vec{H}\vec{C} + \vec{N} + \vec{E}$
 is aggregate complex Gaussian noise that is temporally white
 with $\vec{H}$-conditional spatial correlation 
$\frac{2\alpha\kappa}{N}\vec{HH}\herm + \vec{I} + 
   \beta\big(\frac{2\alpha(1+\kappa)}{N}\diag(\vec{HH}\herm) + \vec{I}\big)$.

 Due to the fact that $\frac{1}{2T}\vec{XX}\herm = \vec{I}$,
 the channel estimate \eqref{Hhat} takes the form
\begin{equation}
   \sqrt{\alpha}\hvec{H}
   = \frac{1}{2T} \vec{YX}\herm
   = \sqrt{\alpha}\vec{H} + \frac{1}{2T}\vec{WX}\herm ,
\end{equation}
 where $\frac{1}{2T}\vec{WX}\herm$ is Gaussian channel estimation error.
 We now analyze the $\vec{H}$-conditional correlations among the 
 elements of the channel estimation error matrix. 
 We begin by noticing
 \begin{align}
\lefteqn{
   \E\bigg\{\bigg[\frac{1}{2T}\vec{WX}\herm\bigg]_{m,p}\bigg[\frac{1}{2T}\vec{WX}\herm\bigg]^*_{n,q}
   	~\bigg|~\vec{H}\bigg\} 
}\nonumber\\
   &= \frac{1}{(2T)^2}
    	\E\bigg\{\sum_{k}[\vec{W}]_{m,k}[\vec{X}]^*_{p,k}
   	\sum_{l}[\vec{X}]_{q,l}[\vec{W}]^*_{n,l}\bigg|\vec{H}\bigg\} \\
   &= \frac{1}{(2T)^2}
   	\sum_{k,l} [\vec{X}]^*_{p,k} [\vec{X}]_{q,l}
   	\E\big\{[\vec{W}]_{m,k} [\vec{W}]^*_{n,l} \biggiv\vec{H}\big\} .
 \end{align}
 To find $\E\big\{[\vec{W}]_{m,k} [\vec{W}]^*_{n,l} \giv\vec{H}\big\}$,
 we recall that 
 \begin{align}
   \E\big\{[\vec{N}]_{m,k} [\vec{N}]^*_{n,l} \biggiv\vec{H}\big\}
   &= \delta_{m-n}\delta_{k-l} \\
   \E\big\{[\vec{C}]_{q,k} [\vec{C}]^*_{p,l} \biggiv\vec{H}\big\}
   &= \frac{2\kappa}{N}\, \delta_{q-p}\delta_{k-l} \\
   \E\big\{[\vec{E}]_{m,k} [\vec{E}]^*_{n,l} \biggiv\vec{H}\big\}
   &= \beta [\vec{\Phi}]_{m,m} \delta_{m-n}\delta_{k-l} ,
 \end{align}
 implying that
 \begin{align}
 	\lefteqn{
   \E\big\{[\vec{W}]_{m,k} [\vec{W}]^*_{n,l} \biggiv\vec{H}\big\}
   } \nonumber \\
   &= \alpha \sum_{q,p} [\vec{H}]_{m,q} [\vec{H}]^*_{n,p}
   	\E\big\{[\vec{C}]_{q,k}[\vec{C}]^*_{p,l}\biggiv\vec{H}\big\}
 	\nonumber\\&\quad
	+ \E\big\{[\vec{N}]_{m,k} [\vec{N}]^*_{n,l} \giv\vec{H}\big\}
	+ \E\big\{[\vec{E}]_{m,k} [\vec{E}]^*_{n,l} \giv\vec{H}\big\} \\
   &= \delta_{k-l} 
   	\bigg(
   	\alpha \frac{2\kappa}{N} \sum_p [\vec{H}]_{m,p} [\vec{H}]^*_{n,p}
	+ (1+\beta [\vec{\Phi}]_{m,m}) \delta_{m-n} \bigg),
\nonumber
 \end{align}
 which implies that
 \begin{align}
   \lefteqn{
   \E\bigg\{\bigg[\frac{1}{2T}\vec{WX}\herm\bigg]_{m,p}\bigg[\frac{1}{2T}\vec{WX}\herm\bigg]^*_{n,q}
   	~\bigg|~\vec{H}\bigg\} 
   }\nonumber\\
   &= \frac{1}{(2T)^2}
   	\sum_{k} [\vec{X}]^*_{p,k} [\vec{X}]_{q,k}
   	\bigg(
   	\alpha \frac{2\kappa}{N} \sum_p [\vec{H}]_{m,p} [\vec{H}]^*_{n,p}
 \nonumber\\&\quad
	+ (1+\beta [\vec{\Phi}]_{m,m}) \delta_{m-n} \bigg) \\
   &= \delta_{p-q} \frac{1}{2T} 
   	\bigg(
   	\alpha \frac{2\kappa}{N} \sum_p [\vec{H}]_{m,p} [\vec{H}]^*_{n,p}
 \nonumber\\ & \quad
   	+ (1+ \beta [\vec{\Phi}]_{m,m}) \delta_{m-n} \bigg),
						\label{eq:white}
 \end{align}
 where the latter expression follows from the fact that
 $\sum_k[\vec{X}]^*_{p,k} [\vec{X}]_{q,k} = 2T\delta_{p-q}$,
 as implied by $\frac{1}{2T}\vec{XX}\herm=\vec{I}$.
 Equation~\eqref{white} implies the estimation error is
 temporally white with $\vec{H}$-conditional spatial correlation
 \begin{align}
   \vec{D} 
   &\defn  \frac{1}{2T}
   	\bigg( \alpha \frac{2\kappa}{N} \vec{HH}\herm + \vec{I}
	+ \beta \diag(\vec{\Phi}) \bigg) \\
   &= \frac{1}{2T} \bigg(
   	\alpha\frac{2\kappa}{N} \vec{HH}\herm + \vec{I} 
\nonumber\\&\quad+
   	\beta\Big(\alpha\frac{2(1+\kappa)}{N}\diag\Big(\vec{HH}\herm\Big) 
		+ \vec{I}\Big) 
   	\bigg) . 				
 \end{align}

 Our final claim is that the channel estimation error 
 $\frac{1}{2T}\vec{WX}\herm$ is
 statistically equivalent to $\vec{D}^{\frac{1}{2}}\tvec{H}$,
 with $\tvec{H}\in\Complex^{M\times N}$ constructed from
 i.i.d $\mc{CN}(0,1)$ entries.
 This can be seen from the following:
 \begin{align}
 \lefteqn{  
   \E\Big\{[\vec{D}^\frac{1}{2}\tvec{H}]_{m,p}
   	[\vec{D}^{\frac{1}{2}}\tvec{H}]^*_{n,q}
   	\Big\} 
 }\nonumber \\
   &= \E\left\{\sum_{k}[\vec{D}^{\frac{1}{2}}]_{m,k}[\tvec{H}]_{k,p}
   	\sum_{l}[\vec{D}^{\frac{1}{2}}]^*_{n,l}[\tvec{H}]^*_{l,q}\right\} \\
   &= \sum_{k,l} [\vec{D}^\frac{1}{2}]_{m,k} [\vec{D}^\frac{1}{2}]^*_{n,l}
   	\E\big\{[\tvec{H}]_{k,p} [\tvec{H}]^*_{l,q} \big\} \\
   &= \delta_{p-q}
   	\sum_{k} [\vec{D}^\frac{1}{2}]_{m,k} [\vec{D}^\frac{1}{2}]^*_{n,k} \\
   &= \delta_{p-q} [\vec{D}]_{m,n} ,
 \end{align}
 where we used the fact that
 $\E\big\{[\tvec{H}]_{k,p} [\tvec{H}]^*_{l,q} \big\}=\delta_{k-l}\delta_{p-q}$.
\section{Interference Cancellation Details}		\label{app:cancellation}
 In this appendix, we characterize the channel-estimate-conditioned
 covariance of the aggregate interference $\vec{v}\R$, 
 whose expression was given in \eqref{v}.

 Recalling that $\hvec{D}\defn\E\{\vec{D}\giv\hvec{H}\}$, we first establish 
 that $\cov\{\vec{D}^\frac{1}{2}\tvec{H}\vec{x}\giv\hvec{H}\}
 = \hvec{D}\tr(\cov(\vec{x}))$,			
 which will be useful in the sequel.
 To show this,
 we examine the $(m,n)^{th}$ element of the covariance matrix:
 \begin{align}
 \lefteqn{
   [\cov\{\vec{D}^\frac{1}{2}\tvec{H}\vec{x}\giv\hvec{H}\}]_{m,n}
 }\nonumber \\
   &= \E\big\{ [\vec{D}^\frac{1}{2}\tvec{H}\vec{x}]_{m} 
   	[\vec{D}^\frac{1}{2}\tvec{H}\vec{x}]_{n}^* \biggiv\hvec{H}\big\} \\
   &= \E\Big\{   
   	\sum_{p,r}[\vec{D}^\frac{1}{2}]_{m,p}[\tvec{H}]_{p,r}[\vec{x}]_{r}
   	\sum_{q,t}[\vec{D}^\frac{1}{2}]_{n,q}^*[\tvec{H}]_{q,t}^*[\vec{x}]_{t}^*
 	\Biggiv\hvec{H}\Big\} \nonumber\\
   &= \sum_{p,r,q,t}  
   	\E\big\{ [\vec{D}^\frac{1}{2}]_{m,p} [\vec{D}^\frac{1}{2}]^*_{n,q}
	\biggiv\hvec{H}\big\}
  \nonumber \\ &\quad \times
 	\underbrace{
   	\E\big\{ [\tvec{H}]_{p,r} [\tvec{H}]^*_{q,t} \big\} 
 	}_{\delta_{p-q}\delta_{r-t}}
 	\E\big\{ [\vec{x}]_{r} [\vec{x}]_{t}^* \big\}  \\
   &= [\hvec{D}]_{m,n} \tr(\cov\{\vec{x}\}) .
 \end{align}
 Rewriting the previous equality in matrix form, we get 
the desired result.
 As a corollary, we note that 
 $\E\{(\vec{D}^\frac{1}{2}\tvec{H})
   	\cov\{\vec{x}\}
	(\vec{D}^\frac{1}{2}\tvec{H})\herm\giv\hvec{H}\}
   = \hvec{D}\tr(\cov\{\vec{x}\})$,
 which will also be useful in the sequel.

 Next we characterize the 
 $(\hvec{H}\sR,\hvec{H}\RR)$-conditional 
 covariance of the receiver distortion $\vec{e}\R$.
 Recalling that 
 $\cov\{\vec{e}\R\}=\beta\diag(\vec{\Phi}\R)$ where
 $\vec{\Phi}\R=\cov\{\vec{u}\R\}$, we have
 $\cov\{\vec{e}\R\giv\hvec{H}\sR,\hvec{H}\RR\}=\beta\diag(\hvec{\Phi}\R)$ where
 $\hvec{\Phi}\R\defn\cov\{\vec{u}\R\giv\hvec{H}\sR,\hvec{H}\RR\}$. 
 Then, given that $\vec{u}\R=\vec{y}\R-\vec{e}\R$ with 
 $\vec{y}\R$ from \eqref{y}, and using the facts that 
 $\cov(\vec{x}\s+\vec{c}\s)=\vec{Q}\s+\kappa\diag(\vec{Q}\s)$ and
 $\cov(\vec{x}\R+\vec{c}\R)=\vec{Q}\R+\kappa\diag(\vec{Q}\R)$,
 we get
 \begin{align}
   \hvec{\Phi}\R
   &= \rho\R\hvec{H}\sR
   	\big(\vec{Q}\s+\kappa\diag(\vec{Q}\s)\big)
	\hvec{H}\sR\herm
 \nonumber\\&\quad
    	+ \E\big\{ (\vec{D}\sR^{\frac{1}{2}}\tvec{H}\sR)
     	\big(\vec{Q}\s+\kappa\diag(\vec{Q}\s)\big)
	(\vec{D}\sR^{\frac{1}{2}}\tvec{H}\sR)\herm \biggiv \hvec{H}\sR \big\}
 \nonumber\\&\quad
       +\eta\R\hvec{H}\RR
   	\big(\vec{Q}\R+\kappa\diag(\vec{Q}\R)\big)
	\hvec{H}\RR\herm
 \nonumber\\&\quad
    	+ \E\big\{ (\vec{D}\RR^{\frac{1}{2}}\tvec{H}\RR)
     	\big(\vec{Q}\R+\kappa\diag(\vec{Q}\R)\big)
 (\vec{D}\RR^{\frac{1}{2}}\tvec{H}\RR)\herm \biggiv\hvec{H}\RR \big\}
 \nonumber\\&\quad
     + \vec{I} \\
   &= \rho\R\hvec{H}\sR
   	\big(\vec{Q}\s+\kappa\diag(\vec{Q}\s)\big)
	\hvec{H}\sR\herm
 \nonumber\\ & \quad
    	+ \hvec{D}\sR \tr(\vec{Q}\s+\kappa\diag(\vec{Q}\s))
 \nonumber\\ & \quad
       +\eta\R\hvec{H}\RR
   	\big(\vec{Q}\R+\kappa\diag(\vec{Q}\R)\big)
	\hvec{H}\RR\herm
\nonumber\\ & \quad
    	+ \hvec{D}\RR \tr(\vec{Q}\R+\kappa\diag(\vec{Q}\R))
     + \vec{I} .
 \end{align}
 Then,
 \begin{align}
   \hvec{\Phi}\R
   &= \rho\R\hvec{H}\sR
   	\big(\vec{Q}\s+\kappa\diag(\vec{Q}\s)\big)
	\hvec{H}\sR\herm
    	+ (1+\kappa)\hvec{D}\sR\tr(\vec{Q}\s) 
 \nonumber\\&\quad
       +\eta\R\hvec{H}\RR
   	\big(\vec{Q}\R+\kappa\diag(\vec{Q}\R)\big)
	\hvec{H}\RR\herm
 \nonumber\\&\quad
    	+ (1+\kappa)\hvec{D}\RR\tr(\vec{Q}\R) 
     + \vec{I} \\
   &\approx 
     \rho\R\hvec{H}\sR \vec{Q}\s \hvec{H}\sR\herm + \hvec{D}\sR\tr(\vec{Q}\s)
 \nonumber\\ & \quad
     + \eta\R\hvec{H}\RR \vec{Q}\R \hvec{H}\RR\herm + \hvec{D}\RR\tr(\vec{Q}\R)
     + \vec{I} ,
 \end{align}
 where, for the approximation, we assumed $\kappa\ll 1$.
 Thus,
 \begin{align}
   \lefteqn{
   \cov\{\vec{e}\R\giv\hvec{H}\sR,\hvec{H}\RR\}
   } \nonumber\\
   &\approx \beta\big(
   	\rho\R\diag(\hvec{H}\sR \vec{Q}\s \hvec{H}\sR\herm)
	+ \hvec{D}\sR\tr(\vec{Q}\s)
 \nonumber\\ & \quad
   	+ \eta\R\diag(\hvec{H}\RR \vec{Q}\R \hvec{H}\RR\herm)
	+ \hvec{D}\RR\tr(\vec{Q}\R)
 	+ \vec{I} \big) .	\quad			\label{eq:Cove}
 \end{align}

 Finally we are ready to characterize $\hvec{\Sigma}\R$, the 
 $(\hvec{H}\sR,\hvec{H}\RR)$-conditional covariance of $\vec{v}\R$. 
 From \eqref{v},
 \begin{align}
   \hvec{\Sigma}\R
   &= \kappa\rho\R\E\big\{\vec{H}\sR\diag(\vec{Q}\s)\vec{H}\sR\herm
   	\biggiv \hvec{H}\sR \big\}
   	+ \hvec{D}\sR\tr(\vec{Q}\s)
 \nonumber\\&\quad
      +\kappa\eta\R\E\big\{\vec{H}\RR\diag(\vec{Q}\R)\vec{H}\RR\herm
   	\biggiv\hvec{H}\RR \big\}
   	+ \hvec{D}\RR\tr(\vec{Q}\R)
  \nonumber\\&\quad
      +\vec{I} + \cov\{\vec{e}\R\giv\hvec{H}\sR,\hvec{H}\RR\} \\
   &= \kappa\rho\R\hvec{H}\sR\diag(\vec{Q}\s)\hvec{H}\sR\herm
   		+\vec{I} + \cov\{\vec{e}\s\giv\hvec{H}\sR,\hvec{H}\RR\}
 \nonumber\\&\quad
       +\kappa\E\big\{(\vec{D}\sR^\frac{1}{2}\tvec{H}\sR)
       		\diag(\vec{Q}\s)
		(\vec{D}\sR^\frac{1}{2}\tvec{H}\sR\herm)\herm
   	\biggiv\hvec{H}\sR \big\}
 \nonumber\\&\quad
   	+ \hvec{D}\sR\tr(\vec{Q}\s)
   	+ \hvec{D}\RR\tr(\vec{Q}\R)
       +\kappa\eta\R\hvec{H}\RR\diag(\vec{Q}\R)\hvec{H}\RR\herm
 \nonumber\\&\quad
       +\kappa\E\big\{(\vec{D}\RR^\frac{1}{2}\tvec{H}\RR)
       		\diag(\vec{Q}\R)
		(\vec{D}\RR^\frac{1}{2}\tvec{H}\RR\herm)\herm
   	\biggiv\hvec{H}\RR \big\} 
  \\
   &= \kappa\rho\R\hvec{H}\sR\diag(\vec{Q}\s)\hvec{H}\sR\herm
       +(1+\kappa)\hvec{D}\sR\tr(\vec{Q}\s)
   \nonumber\\&\quad
       +\kappa\eta\R\hvec{H}\RR\diag(\vec{Q}\R)\hvec{H}\RR\herm
   	+(1+\kappa)\hvec{D}\RR\tr(\vec{Q}\R)
   \nonumber\\&\quad
      +\vec{I} + \cov\{\vec{e}\s\giv\hvec{H}\sR,\hvec{H}\RR\} \\
   &\approx \vec{I} + \kappa\rho\R\hvec{H}\sR\diag(\vec{Q}\s)\hvec{H}\sR\herm
       +\hvec{D}\sR\tr(\vec{Q}\s)
  \nonumber\\&\quad
       +\kappa\eta\R\hvec{H}\RR\diag(\vec{Q}\R)\hvec{H}\RR\herm
   	+\hvec{D}\RR\tr(\vec{Q}\R)
     \nonumber\\&\quad
      +\beta\rho\R\diag(\hvec{H}\sR \vec{Q}\s \hvec{H}\sR\herm)
   	+ \beta\eta\R\diag(\hvec{H}\RR \vec{Q}\R \hvec{H}\RR\herm) ,
 \end{align}
 where, for the approximation, we assumed $\kappa\ll 1$ and $\beta\ll 1$,
 and we leveraged \eqref{Cove}. 
\section{ Gradient Details}				\label{app:grad_proj}

\setcounter{equation}{82}
\begin{figure*}[]
 \normalsize
 \begin{IEEEeqnarray}{rCl}
\lefteqn{
	\frac{\partial }{\partial \vec{Q}\R[l]} 
	\underline{I} (\vec{Q}\s[1], \vec{Q}\s[2], \vec{Q}\R[1],\vec{Q}\R[2],\zeta)
}\nonumber \\
	&=& \frac{\partial }{\partial \vec{Q}\R[l]} \tau[l] 
		\Big\{ (1-\zeta) \big(
			\log\det (  \vec{S}\D[l]  )
			- \log\det ( \hvec{\Sigma}\D[l]  ) \big)
		 + \zeta\big( \log \det (  \vec{S}\R[l]  )
		-  \log \det (  \hvec{\Sigma}\R[l]  )
			\big) \Big\} \\
	&=&	\frac{\partial }{\partial \vec{Q}\R[l]} \tau[l] \bigg\{
			(1-\zeta) \log \det \Big( 
				\rho\D \hvec{H}\RD \vec{Q}\R[l] \hvec{H}\RD\herm
				+ \kappa \rho\D \hvec{H}\RD \diag(\vec{Q}\R[l]) \hvec{H}\RD\herm
				+ \beta \rho\D \diag(\hvec{H}\RD \vec{Q}\R[l] \hvec{H}\RD\herm ) 
				+ \hvec{D}\RD \tr{\vec{Q}\R[l]}
				+ \vec{Z}_1[l]
			\Big) 
\nonumber\\ &&  \quad  
			  - (1-\zeta) \log \det \Big( 
				\kappa \rho\D \hvec{H}\RD \diag(\vec{Q}\R[l]) \hvec{H}\RD\herm
				+ \beta \rho\D \diag(\hvec{H}\RD \vec{Q}\R[l] \hvec{H}\RD\herm )
				+ \hvec{D}\RD \tr{\vec{Q}\R[l]}
				+ \vec{Z}_2[l]
				\Big)
\nonumber \\ &&  \quad
		      + \zeta \log\det \Big(
					\beta\eta\R \diag(\hvec{H}\RD \vec{Q}\R[l] \hvec{H}\RD\herm) 
					+\kappa\eta\R \hvec{H}\RR \diag(\vec{Q}\R[l]) \hvec{H}\RR\herm
					+ \hvec{D}\RR \tr{\vec{Q}\R[l]}
					+\vec{Z}_3[l]
				\Big)
\nonumber \\ && \quad
			  - \zeta \log\det \Big(
					\beta\eta\R \diag(\hvec{H}\RD \vec{Q}\R[l] \hvec{H}\RD\herm)
					+\kappa\eta\R \hvec{H}\RR \diag(\vec{Q}\R[l]) \hvec{H}\RR\herm
					+ \hvec{D}\RR \tr{\vec{Q}\R[l]}
					+\vec{Z}_4[l]
				\Big)
			\bigg\}  \\ 
	&=&	\frac{(1-\zeta)\rho\D}{\ln 2} \tau[l] \Big\{ \big(
			\hvec{H}\RD\herm \big[ 
					\vec{S}\D^{-1}[l]
					+ \beta \diag ( \vec{S}\D^{-1}[l]  - \hvec{\Sigma}\D^{-1}[l])
				\big] \hvec{H}\RD
			\big)\tran 
			+  \kappa \diag \big(  
				\hvec{H}\RD\herm ( \vec{S}\D^{-1}[l]  - \hvec{\Sigma}\D^{-1}[l]) \hvec{H}\RD 
			\big)   \Big\}
\nonumber\\ && \quad
			+ \frac{1-\zeta}{\ln 2} \tau[l] \elsum \big( \hvec{D}\RD \odot ( \vec{S}\D^{-1}[l]  - \hvec{\Sigma}\D^{-1}[l])\tran \big)  \vec{I}  
		+ \frac{\zeta\eta\R}{\ln 2} \tau[l]
		\Big\{
			\kappa \diag \big(
				\hvec{H}\RR\herm ( \vec{S}\R^{-1}[l] - \hvec{\Sigma}\R^{-1}[l] ) \hvec{H}\RR				
			\big)
\nonumber\\ && \quad
			+  \beta \big( 
					\hvec{H}\RD\herm \diag ( \vec{S}\R^{-1}[l] - \hvec{\Sigma}\R^{-1}[l] ) \hvec{H}\RD  
				\big)\tran \Big\}
			+  \frac{\zeta}{\ln 2} \tau[l] \elsum \big( 
					\hvec{D}\RR \odot ( \vec{S}\R^{-1}[l]  - \hvec{\Sigma}\R^{-1}[l] )\tran  
				\big) \vec{I} . 
	\IEEElabel{eq:Gradient}
 \end{IEEEeqnarray}
 \hrulefill
 \vspace*{4pt}
\end{figure*}
\setcounter{equation}{75}
In this appendix, we derive an expression for the gradient 
$\nabla_{\vec{Q}\R[l]} \underline{I}(\QQ, \zeta)$ 
by first deriving an expression for the derivative
$\frac{\partial \underline{I}}{\partial \vec{Q}\R[l]}$ and then using the fact that
$\nabla_{\vec{Q}\R[l]}\underline{I} 
= 2\big(\frac{\partial\underline{I}}{\partial \vec{Q}\R[l]}\big)^*$. 

To do this, we first consider the related problem of computing the 
derivative $\partial\det(\vec{Y})/\partial \vec{X}$, where 
\begin{align}
  \vec{Y} 
  &\defn \vec{C}\diag(\vec{X})\vec{D}  +\diag( \vec{E}\vec{X}\vec{F} ) + \vec{G} \tr(\vec{X}) + \vec{Z},  \label{eq:Ygrad}
\end{align}
and where \eqref{Ygrad} can be written elementwise as 
\begin{align}
	[\vec{Y}]_{i,j}
		&= \sum\limits_{m,n} [\vec{C}]_{i,m} [\vec{X}]_{m,n} [\vec{D}]_{n,j} \delta_{m-n}
			+ [\vec{Z}]_{i,j}
		   \label{eq:Yij}
\\&\quad{}
		  +	\sum\limits_{p,q} [\vec{E}]_{i,p} [\vec{X}]_{p,q} [\vec{F}]_{q,j} \delta_{i-j}
		  + [\vec{G}]_{i,j} \sum\limits_{t} [\vec{X}]_{t,t}
		   .	
\nonumber		
\end{align}
Notice that, for $\vec{V}_{r,s}$ defined as a zero-valued matrix except 
for a unity element at row $r$ and column $s$, 
we have 
\begin{align}
\frac{\partial \det(\vec{Y})}{\partial \vec{X}} 
 &= \sum\limits_{r,s} \vec{V}_{r,s}  
	\frac{\partial \det(\vec{Y})}{\partial [\vec{X}]_{r,s}} \label{eq:deriv1} \\
 &= \sum\limits_{r,s} \vec{V}_{r,s}  
	\sum\limits_{i,j}
	\frac{\partial \det(\vec{Y})}{\partial [\vec{Y}]_{i,j}}
	\frac{\partial [\vec{Y}]_{i,j}}{\partial [\vec{X}]_{r,s}}.
\end{align}
Then, using \eqref{Yij}, we get
\begin{align} 
\lefteqn{	\frac{\partial \det(\vec{Y})}{\partial \vec{X}}  }\nonumber\\
		&=	\sum\limits_{r,s} \vec{V}_{r,s}  
			\sum\limits_{i,j}
				\frac{\partial \det(\vec{Y})}{\partial [\vec{Y}]_{i,j}}
				\Big(  
					[\vec{C}]_{i,r} [\vec{D}]_{s,j} \delta_{r-s}
\nonumber\\&\quad 
					+ [\vec{E}]_{i,r} [\vec{F}]_{s,j} \delta_{i-j}
					+ [\vec{G}]_{i,j} \delta_{r-s}
				\Big)
		\\ 
		&= \diag \left(
				\vec{D}\left(\frac{\partial\det\vec{Y}}{\partial \vec{Y}}\right)\tran \! \vec{C}  
			\right) 
		+ \left(
				\vec{F}\diag\left(\frac{\partial\det\vec{Y}}{\partial \vec{Y}}\right)\tran \! \vec{E}
			\right)\tran 
\nonumber\\&\quad
		+ \elsum \left( 
				\vec{G} \odot \left(\frac{\partial\det\vec{Y}}{\partial \vec{Y}}
			\right) \right)\vec{I}
		\\ 
		&=	\det(\vec{Y}) \Big(			
			\diag\big( \vec{D}\vec{Y}^{-1}\vec{C} \big)
			+ \big( \vec{F} \diag( \vec{Y}^{-1} ) \vec{E} \big)\tran 
\nonumber\\ & \quad
			+  \elsum \big( \vec{G} \odot  (\vec{Y}^{-1})\tran  \big)  \vec{I}  
			\Big)	\label{eq:deriv_end} ,
\end{align}
where, for the last step, we used the fact that 
$\frac{\partial\det(\vec{Y})}{\partial \vec{Y}} = \det (\vec{Y}) ( \vec{Y}^{-1} )\tran \! $. 

Applying \eqref{deriv_end} to \eqref{mutinfo}, we can obtain an expression 
for $\frac{\partial\underline{I}}{\partial\vec{Q}\R[l]}$. 
To do so, we think of $\vec{Z}$ in \eqref{Ygrad} as representing the terms 
in $\underline{I}$ that have zero derivative with respect to $\vec{Q}\R[l]$.
Using $\vec{S}\D[l]$ and $\vec{S}\R[l]$ defined in \eqref{Sd}-\eqref{Sr}, 
and recalling the expression for $\hvec{\Sigma}\D[l]$ in \eqref{sigma}, 
the result is given in \eqref{Gradient}, at the top of the page.

Finally, using 
$\vec{G}\R[l] = 2 \big( \frac{\partial\underline{I}}{\partial \vec{Q}\R[l]} \big)^\ast$,
and leveraging the fact that $\vec{S}\D[l]$, $\vec{S}\R[l]$, $\hvec{\Sigma}\D[l]$, 
and $\hvec{\Sigma}\R[l]$ 
are Hermitian matrices, we get the expression for $\vec{G}\R[l]$ in \eqref{G}.
A similar expression results for $\vec{G}\s[l]$.

\bibliographystyle{ieeetr}
\bibliography{macros_abbrev,stc,books,comm,misc,multicarrier}

\begin{IEEEbiography}[{\includegraphics[width=1in,height=1.25in,clip,keepaspectratio]{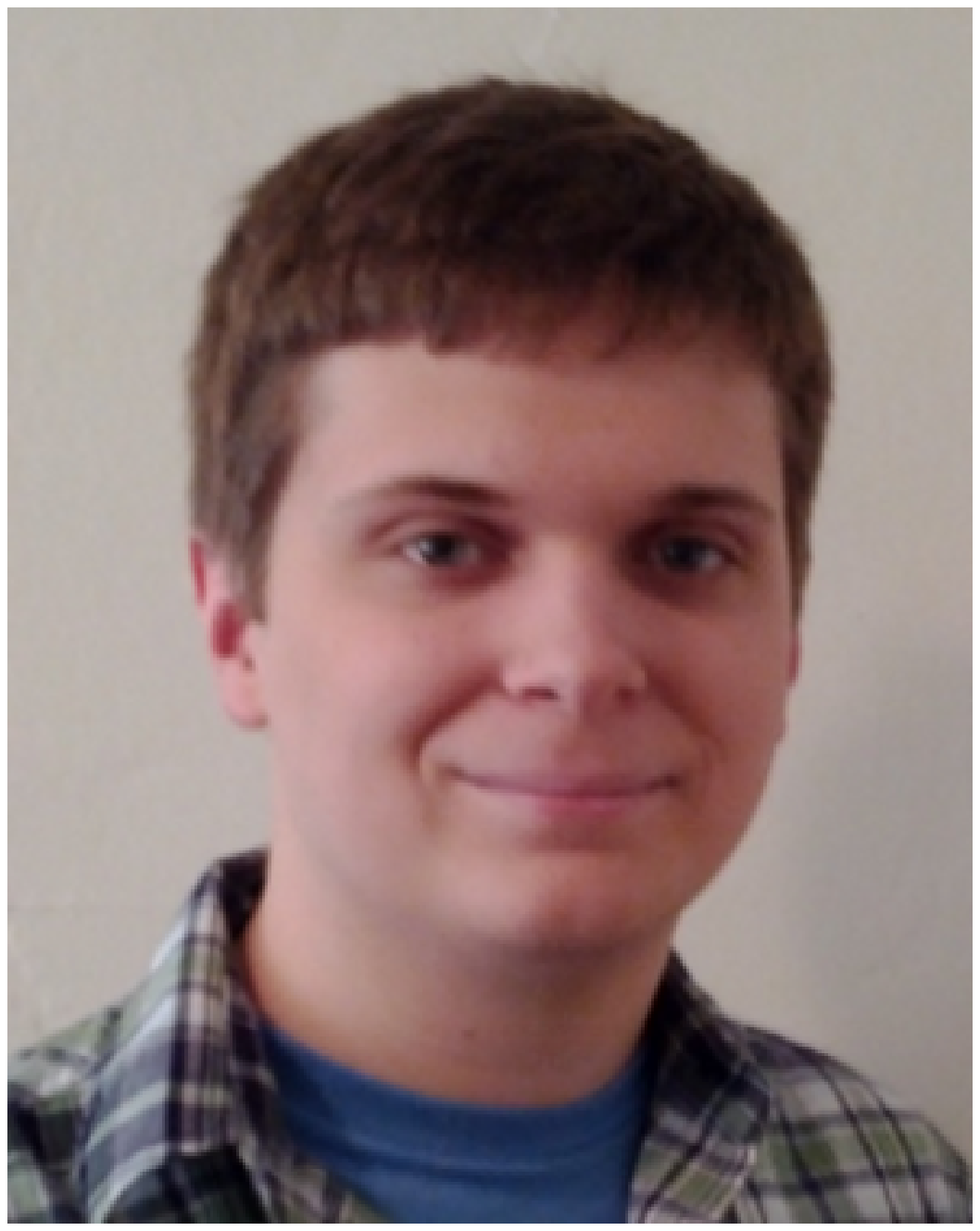}}]{Brian P. Day} received the B.S. in Electrical and Computer Engineering from The Ohio State University in 2010. Since 2010, he has been working toward the Ph.D degree in Electrical and Computer Engineering at The Ohio State University. His primary research interests are full-duplex communication, signal processing, and optimization.
\end{IEEEbiography}

\begin{IEEEbiography}[{\includegraphics[width=1in,height=1.25in,clip,keepaspectratio]{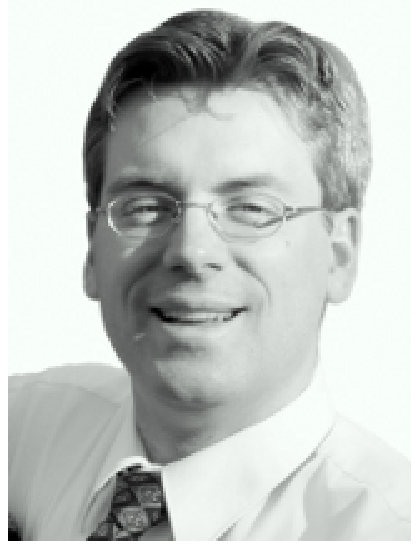}}]{Adam R. Margetts}
received a dual B.S. degree in Electrical Engineering and
Mathematics from Utah State University, Logan, UT in 2000; and the M.S. and
Ph.D. degrees in Electrical Engineering from The Ohio State University,
Columbus, OH in 2002 and 2005, respectively. Dr. Margetts has been with MIT
Lincoln Laboratory, Lexington, MA since 2005 and holds two patents in the
area of signal processing for communications.  His current research
interests include distributed transmit beamforming, cooperative
communications, full-duplex relay systems, space-time coding, and
wireless networking.
\end{IEEEbiography}

\vfill
\pagebreak

\begin{IEEEbiography}[{\includegraphics[width=1in,height=1.25in,clip,keepaspectratio]{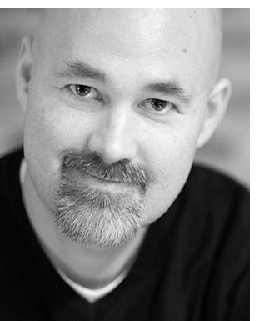}}]{Daniel W. Bliss}
is a senior member of the technical staff at MIT Lincoln
Laboratory in the Advanced Sensor Techniques group.  Since 1997 he has been
employed by MIT Lincoln Laboratory, where he focuses on adaptive signal
processing, parameter estimation bounds, and information theoretic
performance bounds for multisensor systems.  His current research topics
include multiple-input multiple-output (MIMO) wireless communications, MIMO
radar, cognitive radios, radio network performance bounds, geolocation
techniques, channel phenomenology, and signal processing and machine
learning for anticipatory medical monitoring.

Dan received his Ph.D. and M.S. in Physics from the University of California
at San Diego (1997 and 1995), and his BSEE in Electrical Engineering from
Arizona State University (1989).  Employed by General Dynamics (1989-1991),
he designed avionics for the Atlas-Centaur launch vehicle, and performed
research and development of fault-tolerant avionics.  As a member of the
superconducting magnet group at General Dynamics (1991-1993), he performed
magnetic field calculations and optimization for high-energy
particle-accelerator superconducting magnets.  His doctoral work (1993-1997)
was in the area of high-energy particle physics, searching for bound states
of gluons, studying the two-photon production of hadronic final states, and
investigating innovative techniques for lattice-gauge-theory calculations.
\end{IEEEbiography}

\begin{IEEEbiography}[{\includegraphics[width=1in,height=1.25in,clip,keepaspectratio]{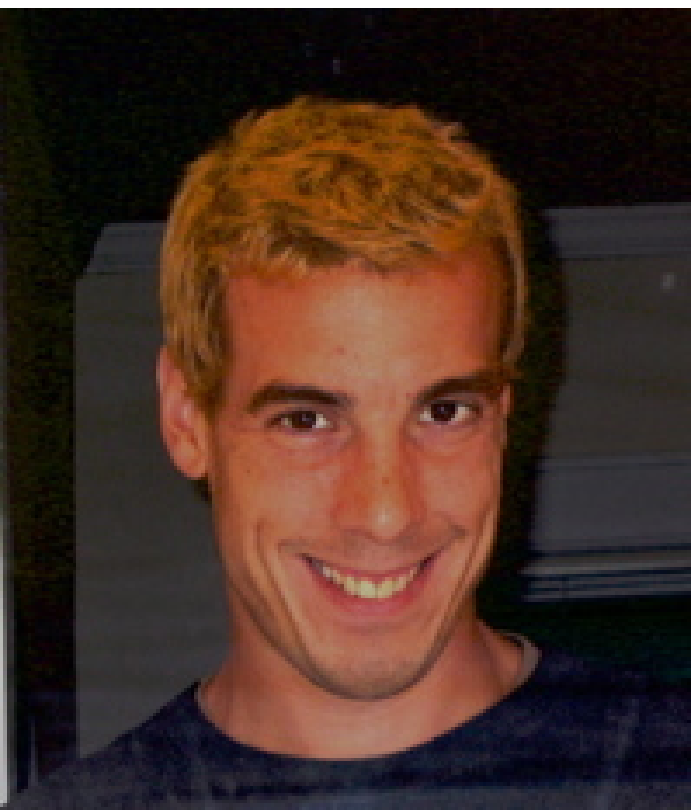}}]{Philip Schniter}
received the B.S. and M.S. degrees in Electrical and Computer Engineering
from the University of Illinois at Urbana-Champaign in 1992 and 1993,
respectively.
From 1993 to 1996 he was employed by Tektronix Inc. in Beaverton, OR as a
systems engineer, and
in 2000, he received the Ph.D. degree in Electrical Engineering from
Cornell University in Ithaca, NY.
Subsequently, he joined the Department of Electrical and Computer Engineering
at The Ohio State University in Columbus, OH, where he is now an Associate
Professor and a member of the Information Processing Systems (IPS) Lab.
In 2003, he received the National Science Foundation CAREER Award, and
in 2008-2009 he was a visiting professor at Eurecom (Sophia Antipolis, France)
and Sup{\'e}lec (Gif-sur-Yvette, France).
Dr. Schniter's areas of interest include statistical signal processing,
wireless communications and networks, and machine learning.
\end{IEEEbiography}

\vfill
\pagebreak


\end{document}